\documentclass[showpacs,superscriptaddress,preprintnumbers,unsortedaddress,notitlepage]{revtex4-1}
\usepackage{graphicx}
\usepackage{dcolumn}
 \usepackage[T1]{fontenc}
\usepackage{mathpazo}
\usepackage{helvet}
\usepackage{amsmath} 
\usepackage{amssymb}
\usepackage{mathtools}
\usepackage{euscript}
\usepackage{fancyhdr} 
\usepackage{wrapfig}
\usepackage{pstricks, pst-node}
\usepackage{subfigure}
\usepackage{units}
\usepackage{slashed}
\usepackage[ngerman,USenglish]{babel}

\begin{document}

\preprint{}

\title{Nonequilibrium chiral fluid dynamics including dissipation and noise}

\author{Marlene Nahrgang}
\affiliation{Institut f\"ur Theoretische Physik, Goethe-Universit\"at and Frankfurt Institute for Advanced Studies (FIAS), Ruth-Moufang-Str.~1, D-60438 Frankfurt am Main, Germany}

\author{Stefan Leupold}
\affiliation{Department of Physics and Astronomy, Uppsala University, Box 516, 75120 Uppsala, Sweden}

\author{Christoph Herold}
\affiliation{Institut f\"ur Theoretische Physik, Goethe-Universit\"at and Frankfurt Institute for Advanced Studies (FIAS), Ruth-Moufang-Str.~1, D-60438 Frankfurt am Main, Germany}

\author{Marcus Bleicher}
\affiliation{Institut f\"ur Theoretische Physik, Goethe-Universit\"at and Frankfurt Institute for Advanced Studies (FIAS), Ruth-Moufang-Str.~1, D-60438 Frankfurt am Main, Germany}

\date{\today}

\begin{abstract}

We present a consistent theoretical approach for the study of nonequilibrium effects in chiral fluid dynamics within the framework of the linear sigma model with constituent quarks. Treating the quarks as an equilibrated heat bath we use the influence functional formalism to obtain a Langevin equation for the sigma field. This allows us to calculate the explicit form of the damping coefficient and the noise correlators. For a selfconsistent derivation of both the dynamics of the sigma field and the quark fluid we have to employ the 2PI (two-particle irreducible) effective action formalism. The energy dissipation from the field to the fluid is treated in the exact formalism of the 2PI effective action where a conserved energy-momentum tensor can be constructed. We derive its form and comment on approximations generating additional terms in the energy-momentum balance of the entire system.
\end{abstract}

 \maketitle

\section{Introduction}
Since neutron stars are too far in space, the Big Bang is too long back in time and astrophysical observations are too indirect, the only possibility to investigate the phase transition of QCD experimentally is by heavy-ion collisions. Here, matter is created under extreme conditions, in small systems and with fast dynamics. It is, therefore, likely that nonequilibrium effects play an important role in the evolution of the fireball. Nonequilibrium methods have also been applied to other systems of extreme conditions, e.g. in the inflationary evolution of the early universe \cite{Starobinsky:1982ee,Boyanovsky:1995ema} and the formation of Bose-Einstein condensates \cite{Boyanovsky:1998yp,Lacaze:2001qf}.


The predictions of most observables of the conjectured critical point are based on the striking feature of diverging fluctuations and correlation lengths in thermodynamic systems at criticality. These should be visible in event-by-event fluctuations of particle multiplicities \cite{Stephanov:1998dy,Stephanov:1999zu}. The proposed experimental measures of fluctuations, which are based on the second moments of the particle distributions did not show any increase in the energy range of $E_{\rm lab}=18-158$ GeV at NA49 \cite{Rybczynski:2008cv}. The key observable for the beam energy scan at RHIC is the net-proton kurtosis \cite{Hatta:2003wn,Stephanov:2008qz}, based on the fourth moment of the distribution. It has the advantage that in thermodynamic systems the kurtosis diverges with higher powers of the correlation length than the second moments \cite{Stephanov:2008qz}. It is, however, unclear if the system created in a heavy-ion collision is in thermal equilibrium at the phase transition. Here, relaxation times become large, whereas the dynamics of a heavy-ion collision is very fast. Even if the system is in equilibrium above the phase transition it is likely to be driven out of equilibrium as it cools through the phase transition. At a second order phase transition this is called critical slowing down and severely limits the growth of the correlation length \cite{Berdnikov:1999ph}. As a consequence, one expects any signal of a critical point to be weakened in a dynamic nonequilibrium situation. Signals of a first order phase transition, however, are based on nonequilibrium effects, such as supercooling. Here, parts of the system remain for some time in the high-temperature phase even below the transition temperature. The decay of this unstable state occurs via either nucleation or spinodal decomposition. A proposed signal for a nonequilibrium situation at a chiral phase transition is the enhancement of soft pions from the decay of disoriented chiral condensates \cite{Bjorken:1991xr,Rajagopal:1993ah,Randrup:1996ay,Mishustin:1998eq,Chomaz:2003dz}.

Nonequilibrium phenomena can be studied in the framework of Langevin dynamics. Here, the chiral fields are split into hard and soft modes. The hard modes constitute a heat bath for the soft modes. Due to the interaction with the hard modes the soft modes undergo additional dissipative processes. Calculations of the influence functional for a system of soft modes interacting with a heat bath of hard modes have extensively been studied in $\phi^4$ theory \cite{Morikawa:1986rp,Gleiser:1993ea,Boyanovsky:1996xx,Greiner:1996dx}, in gauge theories \cite{Bodeker:1995pp,Son:1997qj} and in ${\cal O}(N)$ chiral models \cite{Rischke:1998qy}. However, the back reaction of the soft modes on the hard modes, e.g. the exchange of energy is typically not included in these influence functional studies.

A framework which is based on the conservation of energy and particle numbers is fluid dynamics.
Chiral fluid dynamics combines the dynamics of heavy-ion collisions with the explicit propagation of the chiral fields at the chiral phase transition. The common idea is to embed a low energy effective model of QCD into a fluid dynamic simulation of a heavy-ion collision.
A model that is particularly well suited for this combined approach is the linear sigma model with constituent quarks \cite{GellMann:1960np}. The dynamics of the quarks is reduced to a fluid dynamic evolution of densities. This gives rise to the name chiral fluid dynamics. The quarks are thus assumed to be locally equilibrated and act as a heat bath with local temperatures. In \cite{Mishustin:1998yc} an additional dilaton field was included in the explicit dynamics to model scale invariance breaking in QCD. Strong nonlinear oscillations were found for all the classical fields at the phase transition. From these oscillations the particle production of sigmas, pions and glueballs was calculated. In \cite{Paech:2003fe} initial fluctuations were propagated deterministically through the phase transition. 
The linear sigma model with constituent quarks treats the quarks and antiquarks and the mesons on equal footing. In the real world confining forces recombine quarks and antiquarks in mesons and baryons below the confinement critical temperature. The aspect of confinement is not included in the linear sigma model with constituent quarks. In extensions of the model gluons are included on the level of the dilaton field \cite{Mishustin:1998yc} or the Polyakov loop \cite{Schaefer:2007pw,Stokic:2008jh}.
In all the existing approaches so far \cite{Mishustin:1998yc,Scavenius:1999zc,Paech:2003fe} the time evolution of the chiral fields is given by the classical equations of motion. Thus, the chiral fields are explicitly propagated out of equilibrium with the quarks and do not relax to their equilibrium value for the temperatures given by the heat bath. Relaxational and stochastic processes have been neglected.

A formalism which is capable of including dissipation and noise together with a consistent back reaction on the heat bath is provided by the so-called 2PI formalism. Two-particle irreducible (2PI) diagrams are closed diagrams which do not fall apart when cutting two lines. Consequently the 2PI formalism deals with full one-point (mean-field) and full two-point functions (full propagators). It has been originally developed as a functional formalism in \cite{Luttinger:1960ua} and was extended to yield transport equations in a quantum theory that preserve the macroscopic conservation laws and are thermodynamically consistent in the equilibrium limit \cite{Lee:1960zza,Baym:1961zz,Baym:1962sx}. These can be found from a certain class of approximations, called $\Phi$-derivable, to the Schwinger-Dyson equation for the two-point function.
In the path integral formulation of this approach \cite{Cornwall:1974vz} it turns out that the $\Phi$-functional is identical to the nontrivial part of the two-particle irreducible effective action. In \cite{Ivanov:1998nv} this selfconsistent approach is generalized to arbitrary nonequilibrium many-body systems. Renormalization of $\Phi$-derivable approximations is studied in \cite{vanHees:2001ik,vanHees:2001pf,vanHees:2002bv}.

In this paper we extend existing chiral fluid dynamic models by the consistent inclusion of damping and noise in the dynamics of the order parameter of chiral symmetry, the sigma field. To achieve this we apply two methods of nonequilibrium quantum field theory, the influence functional and the two-particle irreducible effective (2PI) action, to the linear sigma model with constituent quarks. In the influence functional method the separation of the systems seems obvious in our approach to chiral fluid dynamics. We consider the quarks as the heat bath and treat the sigma field as the relevant sector. The influence functional is obtained by perturbatively integrating out the quarks. We obtain the proper Langevin equation for the sigma field. However, in the influence functional method it is not possible to control the local equilibrium properties of the quarks without further assumptions. We need to go beyond existing studies of Langevin equations by putting special emphasis on the properties and the evolution of the heat bath itself. We expect that the back reaction to the heat bath induced by the dynamics of the chiral fields can be important for the overall evolution of the system.

In order to derive the nonequilibrium propagation of the chiral fields and the thermodynamic properties of the quarks selfconsistently we apply the formalism of the 2PI effective action. The analysis of the chiral order parameter, the $\sigma$ field will be restricted to the mean field. One often defines the mean field as an average over quantum and thermal fluctuations in which case there is neither damping nor noise. Within the 2PI effective action formalism the mean field is obtained from an integration over quantum fluctuations only and thus still contains the necessary information about dissipation and noise.

For a first qualitative analysis we restrict our model on the effect of the chiral phase transition and do not include the Polyakov loop. The Polyakov loop extended model describes confinement well on a statistical level and thermodynamic quantities have been calculated \cite{Schaefer:2007pw,Stokic:2008jh} but it is not evident how to treat the Polyakov loop dynamically. 
In our present analysis we also neglect the pion degrees of freedom. Their vacuum expectation value vanishes anyway. From a conceptual point of view their inclusion provides no further problems. Their physical significance is largest below the chiral phase transition. As quasi-Goldstone bosons the pions are very light. Therefore they can be produced from the decay of the sigma meson, leading to additional damping of the sigma field. The hard modes yield a sizable contribution to the heat bath with their own equation of state. Thus this needs to be included in a two-fluid dynamic description. Obtaining the Langevin equation for the soft pion modes is also a straightforward task. The inclusion of pions is devoted to future work. In the present work we explain the formalism of the
coupled system of the explicitly propagated degrees of freedom and the fluid dynamic expansion of the heat bath.
In an upcoming paper we will present detailed numerical results in a one-fluid dynamic set up. For the time being
we concentrate on the heat bath of quarks and antiquarks.

This paper is organized as follows. In section \ref{sec:linsigmamodel} we briefly review the linear sigma model with constituent quarks and its phase structure. The influence functional is calculated in section \ref{sec:siflinsig} including the damping and the noise kernel. In section \ref{sec:2pieacal} we apply the formalism of the 2PI effective action. Both approaches yield the same Langevin equation for the sigma field, calculated in section \ref{sec:sc_eomsigmafield}. In the exact formalism of the 2PI effective action a conserved energy-momentum tensor of the entire system can be constructed. We will derive this form and comment on the energy-momentum conservation of approximations to the full equations in section \ref{sec:sc_energyconservation}.

\section{The linear sigma model with constituent quarks}\label{sec:linsigmamodel}

The linear sigma model with nucleons \cite{GellMann:1960np} has been studied for years as the prototype effective model of dynamic chiral symmetry breaking. The coupling to dynamic quark degrees of freedom instead of nucleons yields the additional feature that the quarks with light current masses obtain a heavy mass at the phase transition and thus turn into constituent quarks \cite{Jungnickel:1995fp,Berges:1998ha,Tetradis:2003qa,Schaefer:2006sr}. 
The Lagrangian reads 
\begin{equation}
{\cal L}=\bar{q}[i\gamma^\mu\partial_\mu-g(\sigma+i\gamma_5\vec\tau{\vec\pi})]q 
  + \frac{1}{2}(\partial_\mu\sigma\partial^\mu\sigma)+
 \frac{1}{2}(\partial_\mu\vec{\pi}\partial^\mu\vec{\pi}) 
- U(\sigma, \vec{\pi}) \, ,
\label{eq:LGML}
\end{equation}
where $q=\left(u,d\right)$ is the constituent quark field, $\sigma$ the sigma field and $\vec{\pi}$ the pionic fields. The strength of the coupling between the quarks and the chiral fields is $g$. In the vertex for the pion-quark coupling the $\gamma_5$ matrix enters to account for the pseudoscalar nature of the $\pi$ mesons and the isospin Pauli matrices $\vec{\tau}$ for the isospin degeneracy of the pions. The interaction between the chiral fields is given by the potential 
\begin{equation}
U\left(\sigma, \vec{\pi}\right)=\frac{\lambda^2}{4}\left(\sigma^2+\vec{\pi}^2-\nu^2\right)^2-h_q\sigma-U_0\, .
\label{eq:Uchi}
\end{equation} 
The Lagrangian (\ref{eq:LGML}) is invariant under ${\rm SU}(2)_{\rm L}\times{\rm SU}(2)_{\rm R}$ symmetry transformations if the explicit symmetry breaking term $h_q$ vanishes in the potential (\ref{eq:Uchi}).
The parameters in (\ref{eq:Uchi}) are chosen such that chiral symmetry is spontaneously broken in the vacuum, where $\langle\sigma\rangle=f_\pi=93$~MeV and $\langle\vec\pi\rangle=0$. The explicit symmetry breaking term taking into account the finite quark masses is $h_q=f_\pi m_\pi^2$ with the pion mass $m_\pi=138$~MeV. With these requirements, $\nu^2=f_\pi^2-m_\pi^2/\lambda^2$. Choosing $\lambda^2=20$ yields a realistic vacuum sigma mass $m_\sigma^2=2\lambda^2 f_\pi^2 + m_\pi^2\approx 604$~MeV. In order to have zero potential energy in the ground state the term $U_0=m_\pi^4/(4\lambda^2)-f_\pi^2 m_\pi^2$ is subtracted. At a coupling $g=3.3$ the constituent quark mass in vacuum is $m_q=306.9$~MeV.


The one-loop thermodynamic potential in mean-field approximation is
\begin{equation}
 \Omega(T,\mu)=-\frac{T}{V}\ln Z=U\left(\sigma, \vec{\pi}\right)+\Omega_{q\bar q}\, ,
\label{eq:lsm_omegaq0}
\end{equation}
with the chiral potential $U(\sigma,\vec\pi)$ and the quark contribution
\begin{equation} 
\Omega_{q\bar q}(T,\mu)=-d_q \int\frac{{\rm d}^3p}{(2\pi)^3}\left(E+T\ln\left(1+\exp\left(\frac{\mu-E}{T}\right)\right)+T\ln\left(1+\exp\left(\frac{-\mu-E}{T}\right)\right)\right)\, ,
 \label{eq:lsm_omegaq1}
\end{equation}
where $\mu$ is the quark chemical potential and  $d_q=12$ is the degeneracy factor of the quarks for $N_f=2$ flavors, $N_c=3$ colors and the two spin states. Though technically very similar there is a noteworthy difference to the calculation of the partition function for free fermions with mass $m_f$. In the Lagrangian (\ref{eq:LGML}) there is no fermionic mass. The quark mass is generated by nonvanishing expectation values of the chiral fields due to spontaneous symmetry breaking. During the evaluation of the functional determinant in Dirac and isospin space one generates a term defined as the effective mass of the quarks
\begin{equation}
 m_{\rm eff}^2=g^2(\sigma^2+\vec\pi^2)\, .
\end{equation}
Then, the energy of the quarks and antiquarks is
\begin{equation}
 E=\sqrt{\vec p^2+m_{\rm eff}^2}=\sqrt{\vec p^2+g^2(\sigma^2+\vec\pi^2)}\, .
\end{equation}
This is obviously not a medium-independent quantity as the chiral expectation values depend on both medium parameters $T$ and $\mu$. Thus, the divergent term in (\ref{eq:lsm_omegaq1}) cannot be subtracted as a simple zero-temperature contribution. It needs to be renormalized more carefully. By using standard renormalization techniques one part of the divergence can be absorbed in the parameters $\lambda$ and $\nu$ of the classical potential $U(\sigma,\vec\pi)$, while a logarithmic term depending on $m_{\rm eff}$ and the renormalization scale remains.
In \cite{Skokov:2010sf} it was shown that by neglecting this contribution one fails to reproduce the second order phase transition for $\mu=0$ in the chiral limit. In \cite{Mocsy:2004ab} the renormalization scale dependence was investigated phenomenologically. A thorough study of medium dependent corrections to mean-field calculations, perturbative and renormalization group approaches to Yukawa theory, also shows a crucial effect on the phase structure \cite{Palhares:2008yq,Fraga:2009pi,Palhares:2010be}.

To achieve the goal of this work, namely the coupling of chiral nonequilibrium dynamics at the phase transition to a fluid dynamic expansion of the matter, we need a field-theoretical model exhibiting a phase transition. This is given by the mean-field approximation \cite{Scavenius:2000qd} and we can neglect the effects of the vacuum correction.

The pressure of the system is
\begin{equation}
  p(T,\mu)=-\Omega(T,\mu)\, ,
\label{eq:pressure}
\end{equation}
from which all thermodynamic quantities can be calculated. We are especially interested in the energy density. It is given by the thermodynamic relation
\begin{equation}
 e(T,\mu)= Ts-p+\mu n\, ,
\end{equation}
with the entropy density $s=(\partial p/\partial T)_\mu$ and the baryon density $n=-(\partial p/\partial \mu)_T$. Then
\begin{equation}
 e(T,\mu)= T\left(\frac{\partial p(T,\mu)}{\partial T}\right)\bigg|_\mu-p(T,\mu)-\mu\left(\frac{\partial p(T,\mu)}{\partial \mu}\right)\bigg|_T\, .
\label{eq:chfl1_energydens}
\end{equation}

\begin{figure}
 \centering
 \includegraphics{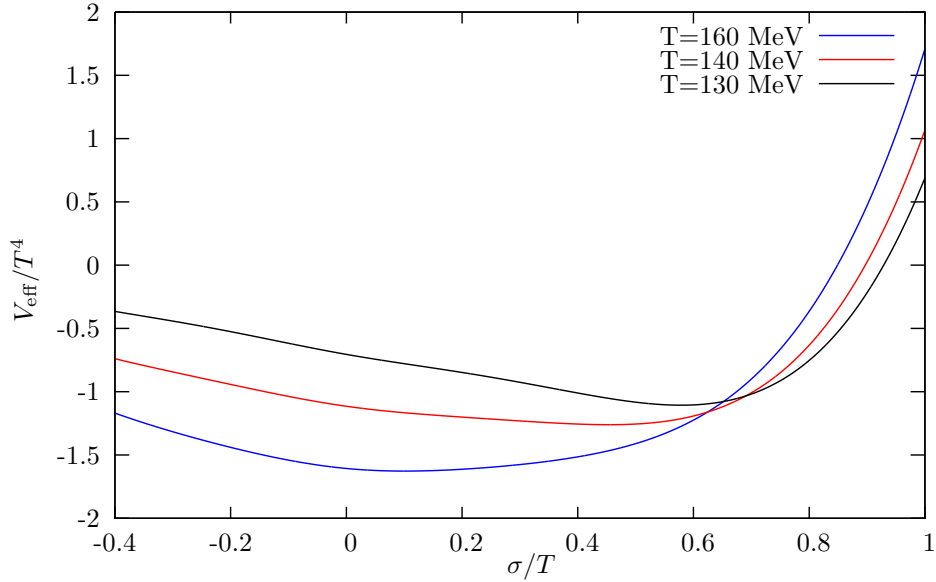}
\caption[The effective potential at $g=3.3$.]{The effective potential for a coupling $g=3.3$ and three different temperatures. The potential changes smoothly between the high and the low temperature phase. It describes a generic crossover transition.}
\label{fig:chfl1_Veffco}
\end{figure}

\begin{figure}
 \centering
 \includegraphics{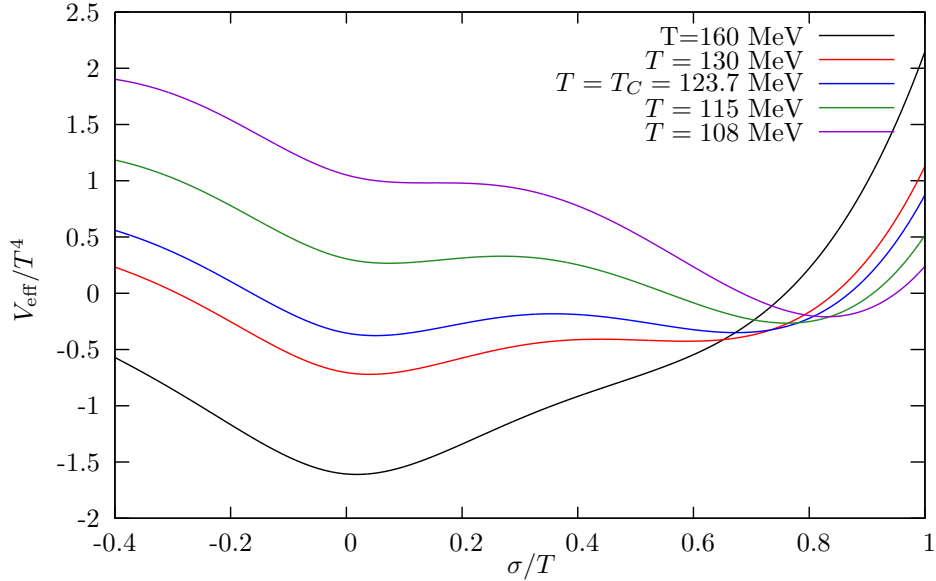}
\caption[The effective potential at $g=5.5$.]{The effective potential for a coupling $g=5.5$ and temperatures $T>T_c$, $T<T_c$ and the critical temperature. At the critical temperature the two minima are degenerate and represent the two coexisting phases for a first order phase transition.}
\label{fig:chfl1_Vefffo}
\end{figure}

The current understanding of the phase diagram of QCD is based on lattice results, which reliably predict that the phase transition is a crossover for zero baryochemical potential \cite{Aoki:2006we}, and on less settled model studies, which claim that at larger baryochemical potentials and lower temperatures the phase transition is of first order \cite{Friman:2011zz}. Then, this first order phase transition ends in a critical point. The phase structure of the linear sigma model with constituent quarks in mean-field approximation is qualitatively the same. For a first qualitative analysis, we can fix the baryochemical potential at $\mu=0$ and tune the strength of the phase transition by changing the coupling constant $g$ \cite{Scavenius:2000bb,Aguiar:2003pp}. At $\mu=0$ the thermodynamic potential is
\begin{equation}
  \Omega(T,\phi)=U\left(\sigma, \vec{\pi}\right)+\Omega_{q\bar q}
		=U\left(\sigma, \vec{\pi}\right)-2d_q T \int\frac{{\rm d}^3p}{(2\pi)^3}\log\left(1+\exp\left(-\frac{E}{T}\right)\right)\, .
\end{equation}
The sigma mass is given by the curvature of the thermodynamic potential at the equilibrium values of the chiral fields
\begin{equation}
 m_\sigma^2=\frac{\partial^2\Omega}{\partial\sigma^2}|_{\sigma=\sigma_{\rm eq}}\, .
\end{equation}
For the realistic coupling, $g=3.3$, and $\mu=0$ the effective potential changes smoothly from the high-temperature phase to the low-temperature phase, see figure (\ref{fig:chfl1_Veffco}). For higher couplings $g$ the effective potential starts to exhibit a first order phase transition. In figure (\ref{fig:chfl1_Vefffo}) we show the effective potential for various temperatures and $g=5.5$. 

Above the critical temperature there is one minimum close to $\sigma\simeq0$, which is the minimum without explicit symmetry breaking. At a critical temperature of $T_c=123.27$~MeV the two minima are degenerate and form the two coexisting phases. Below $T_c$ the high-temperature minimum becomes unstable but exists down to the spinodal temperature $T_{\rm sp}=108$~MeV \cite{Scavenius:2000bb}.

\begin{figure}
 \centering
 \includegraphics{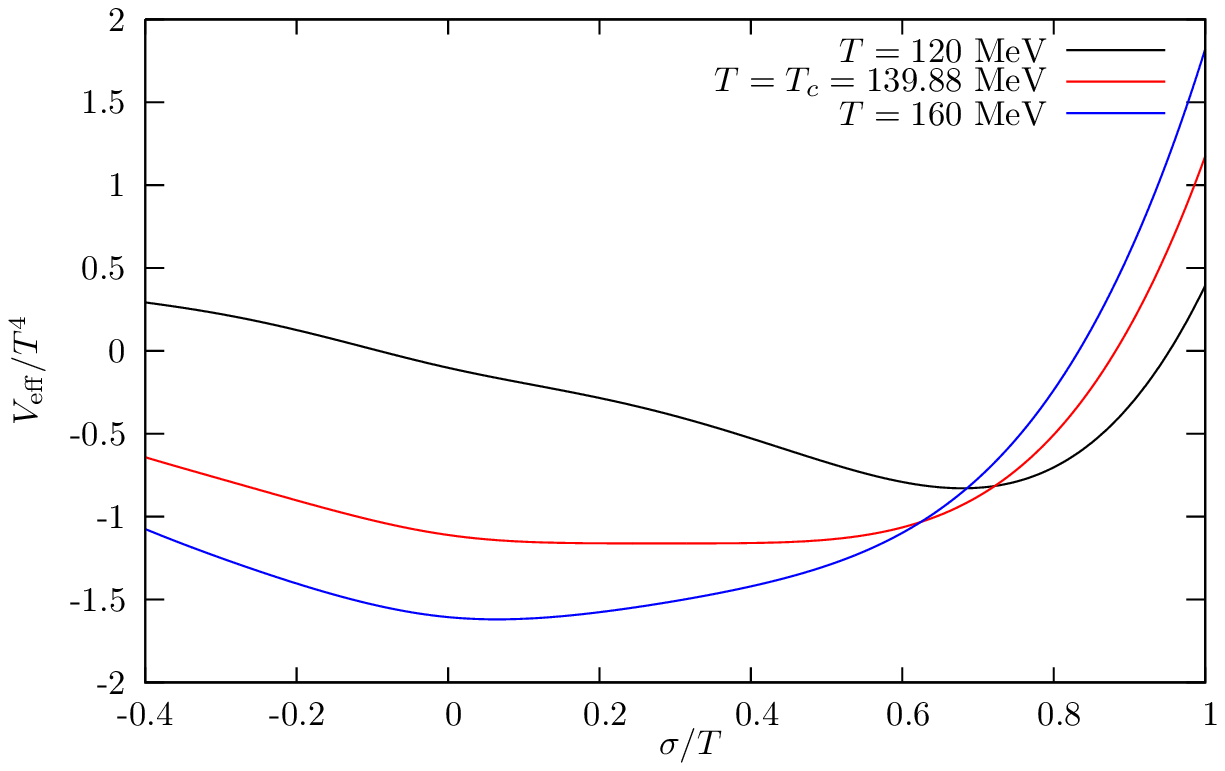}
\caption[The effective potential at $g=3.63$.]{The effective potential for a temperature above, a temperature below and the critical temperature at a coupling $g=3.63$. At the critical temperature the minimum becomes very flat. This indicates a second order phase transition.}
\label{fig:chfl1_Veffcp}
\end{figure} 
If one carefully looks for the vanishing of the barrier by decreasing the coupling $g$, the effective potential shows the shape of a second order phase transition for $g=3.63$ with a critical temperature of $T_c=139.88$~MeV. Here, the curvature at the minimum becomes very flat, see figure \ref{fig:chfl1_Veffcp}, and the sigma mass decreases to $m_\sigma=26.6$~MeV. This value of $m_\sigma$ is significantly smaller than the vacuum mass, but still not zero. In order to lower the sigma mass further it would be necessary to tune the parameters $g$ and $T_c$ more precisely. 

\begin{figure}
 \centering
 \includegraphics{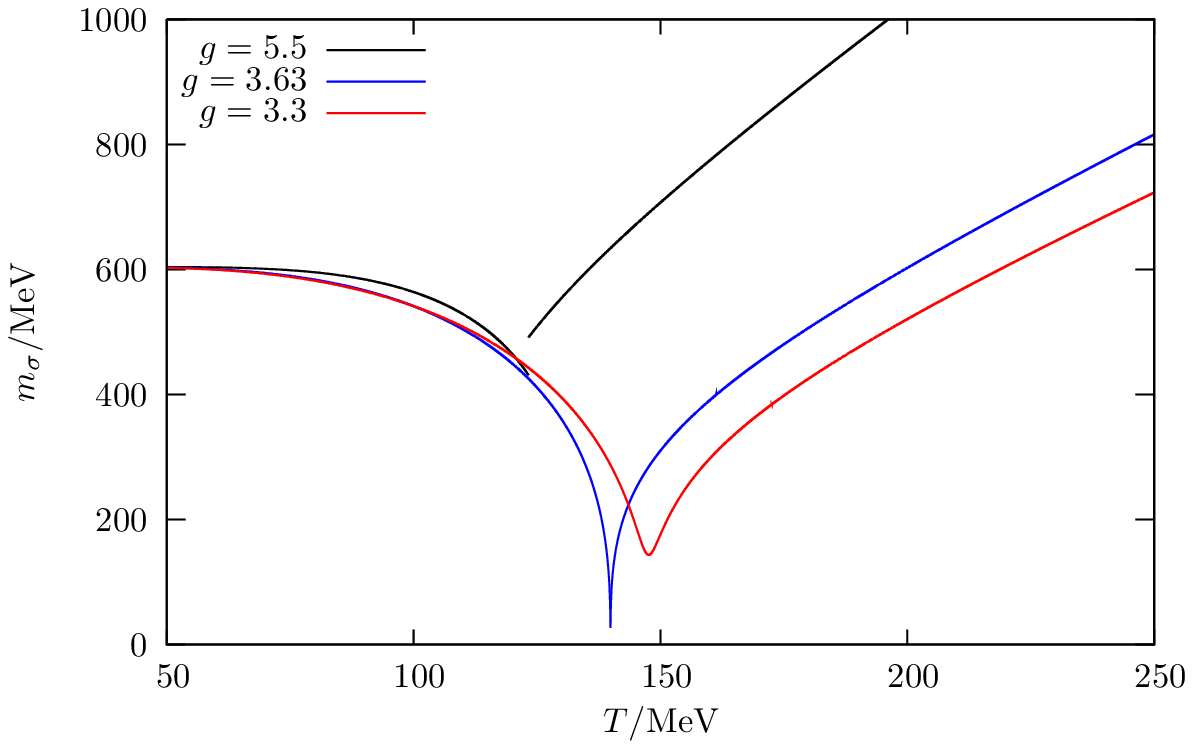}
 \caption[The equilibrium value of the sigma mass $m_\sigma$.]{The equilibrium value of the sigma mass $m_\sigma$ for the three scenarios: with a first order phase transition $g=5.5$, a critical point $g=3.63$ and a crossover $g=3.3$. We also observe a discontinuity in $m_\sigma$ at the first order phase transition.}
 \label{fig:chfl1_sigmamasseq}
\end{figure}

For all three phase transition scenarios we show the temperature dependence of the mass of the sigma field in equilibrium $m_\sigma$ in figure \ref{fig:chfl1_sigmamasseq}.

\section{The influence functional for the linear sigma model with constituent quarks}\label{sec:siflinsig}
The influence functional method \cite{Feynman:1963fq,Greiner:1998vd} gives a reduced description of the entire system with focus on the evolution of the relevant variables. The details of the environment are eliminated by integrating out the environmental fields in a path integral over the closed time path contour \cite{Schwinger:1960qe,Keldysh:1964ud}. 
For the linear sigma model with constituent quarks we assume the following splitting: the irrelevant degrees of freedom are the quarks and antiquarks, which constitute the environment, or heat bath, and which are propagated fluid dynamically on the level of densities, and the relevant sector is that of the chiral fields, which we propagate explicitly. 
Again, we keep the pion fields fixed at their vacuum expectation value $\langle\pi\rangle=0$. The free action of the quarks and antiquarks reads
\begin{equation}
 S_0[q,\bar q]=i\int{\rm d}^4x\bar q(x)\gamma^\mu\partial_\mu q(x)\, ,
\end{equation}
and the interaction between the quarks and the sigma field is of Yukawa type
\begin{equation}
 S_{\rm int}[q,\bar{q},\sigma]=-g\int{\rm d}^4x\bar q(x) q(x)\sigma(x)\, .\\
\end{equation}


The influence functional is calculated from a perturbative expansion in g. This might seem doubtful because $g$ is of order ${\cal O}(1)$. It must, however, be looked at the individual processes connected with the orders of the expansion. Higher orders in $g$ involve more sigma modes and quark-antiquark pairs. The lower the density of the system the less likely these processes become and, thus, contribute less to the damping of the sigma field. Still, this is a crucial point in the application of the influence functional method. The explicit calculation of the influence functional to second order, which is the first nontrivial order, using standard techniques on the Keldysh-contour is deferred to appendix \ref{sec:noneqsif}. It is convenient to write the influence functional in center $\bar\sigma$ and relative variable $\Delta\sigma$ of the upper ($+$) and the lower ($-$) branch of the contour.
It has one term that is linear and one term that is quadratic in $\Delta\sigma$,
 \begin{equation}
   iS_{\rm IF}[\bar\sigma,\Delta\sigma]= i\int\!\!{\rm d}^4xD(x)\Delta\sigma(x)-\frac{1}{2}\int\!\!{\rm d}^4x\int\!\!{\rm d}^4y\Delta\sigma(x){\cal N}(x,y)\Delta\sigma(y)\, ,
 \label{eq:sc_siflinsig5}
 \end{equation}
with the damping kernel
\begin{equation}
 D(x)=ig^2\int_{y_0}^{x_0}\!\!{\rm d}^4y\bar\sigma(y)\biggl(S^<(x-y)S^>(y-x)-S^>(x-y)S^<(y-x)\biggr)\, ,
\label{eq:sc_dampingkernel}
\end{equation}
and the noise kernel
\begin{equation}
 {\cal N}(x,y)=-\frac{1}{2}g^2\biggl(S^<(x-y)S^>(y-x)+S^>(x-y)S^<(y-x)\biggr)\, .
\label{eq:sc_noisekernel}
\end{equation}
The influence functional
\begin{equation}
 S_{\rm IF}[\bar\sigma,\Delta\sigma]
   =\int\!\!{\rm d}^4xD(x)\Delta\sigma(x)+\frac{i}{2}\int\!\!{\rm d}^4x\int\!\!{\rm d}^4y\Delta\sigma(x){\cal N}(x,y)\Delta\sigma(y)\, ,
\end{equation}
has an imaginary part. It is exactly this term that causes the underlying quantum system to decohere and allows for a classical description of the system's dynamics. This means that every trajectory can be assigned a unique probability. For these trajectories the relative field variable vanishes $\Delta\sigma(x)=0$, because trajectories that have a significantly large $\Delta\sigma$ are exponentially suppressed. Together with the coarse graining of the environment, the decoherence leads to the quantum-to-classical transition of the system. Fluctuations in the classical equation of motion appear as a remnant of coarse graining and decoherence \cite{Hu:1991di,Hu:1993vs,GellMann:1992kh,Calzetta:1995ys,Calzetta:1995ea}.

The zeroth order in the perturbative calculation of the influence functional gives one due to normalization and the first order vanishes. Then, the second order contribution $S_{\rm IF}^{(2)}$ is equal to $i$-times the influence functional itself, as can be seen by taking the logarithm of (\ref{eq:sc_siflinsig})
\begin{equation}
 iS_{\rm IF}[\sigma^{\scriptscriptstyle +},\sigma^{\scriptscriptstyle -}]=\ln(1+S_{\rm IF}^{(2)}[\sigma^+,\sigma^-])\simeq S_{\rm IF}^{(2)}[\sigma^{\scriptscriptstyle +},\sigma^{\scriptscriptstyle -}]\, .
\end{equation}
We also note that in the perturative approach to the influence functional we neither obtain the thermal mass correction nor the equilibrium properties of the heat bath. For the calculation of the mass correction one needs to include further information. For example, it is possible to find this term by directly calculating the equation of motion as it was done in \cite{Rischke:1998qy} for $\phi^4$ theory. In the formalism of the 2PI effective action this is included automatically.

\subsection{The noise kernel and fluctuations}\label{sec:noisekernel}
At first glance, the noise kernel seems to be redundant, because it is quadratic in $\Delta\sigma(x)$ and, therefore, vanishes after variation with respect to $\Delta\sigma(x)$. 
However, the semiclassical concept of obtaining the equation of motion from varying the action is well-defined only for real actions. Here, the noise kernel introduces an imaginary part, which we need to rewrite in order to obtain a real action. This is done by introducing a new stochastic field $\xi$, which discloses the physical significance of the noise kernel. The imaginary part of $S_{\rm IF}$ can be rewritten by making use of the Gauss integral evaluation
\begin{equation}
 \exp\left(-\frac{1}{2}\int\!\!{\rm d}^4x\int\!\!{\rm d}^4y\Delta\sigma(x){\cal N}(x,y)\Delta\sigma(y)\right)=\int{\cal D}\xi P[\xi] \exp\left(i\int\!\!{\rm d}^4x\xi(x)\Delta\sigma(x)\right)\, .
\end{equation}
Here, the stochastic weight $P[\xi]$ is a Gauss distribution
\begin{equation}
 P[\xi]=N'\exp\left(-\frac{1}{2}\int\!\!{\rm d}^4x\int\!\!{\rm d}^4y\xi(x){\cal N}^{-1}(x,y)\xi(y)\right)\, ,
\label{eq:sc_gaussdissxi}
\end{equation}
with a normalization constant $N'$. Then, the stochastic field $\xi$ is fully determined by its first two moments, a vanishing expectation value and the variance:
\begin{subequations}
  \begin{align}
 \langle\xi(x)\rangle&=0\, ,\\
 \langle\xi(x)\xi(y)\rangle&={\cal N}(x,y)\, .
\end{align}
\end{subequations}
This stochastic force $\xi$ plays an essential role in the equilibration of the classical fields. By the dissipation-fluctuation theorem it enforces the relaxation to the correct equilibrium state \cite{einstein}.

\subsection{The semiclassical equations of motion}
The semiclassical equations of motion for the sigma field are obtained from the stochastic influence functional $\tilde S_{\rm IF}$, defined in
\begin{equation}
 \exp(iS_{\rm IF})=\int{\cal D}\xi P[\xi]\exp\left(i\int\!\!{\rm d}^4x(D(x)+\xi(x))\Delta\sigma(x)\right)=\int{\cal D}\xi P[\xi]\exp\left(i\tilde S_{\rm IF}\right)\, ,
\end{equation}
by varying
\begin{equation}
 S_{\rm cl}[\sigma^{\scriptscriptstyle+}]-S_{\rm cl}[\sigma^{\scriptscriptstyle-}]+\tilde S_{\rm IF}[\bar\sigma,\Delta\sigma]
\end{equation}
with respect to $\Delta\sigma$ and then setting $\Delta\sigma=0$. From the classical action one obtains
\begin{equation}
 \frac{\delta (S_{\rm cl}[\sigma^{\scriptscriptstyle+}]-S_{\rm cl}[\sigma^{\scriptscriptstyle-}])}{\delta\Delta\sigma}\bigg|_{\Delta\sigma=0}=\frac{\delta S_{\rm cl}[\bar\sigma]}{\delta\bar\sigma}\, .
\end{equation}
 The semiclassical Langevin equation for the sigma field is
\begin{equation}
 -\frac{\delta S_{\rm cl}[\bar\sigma]}{\delta\bar\sigma}-D=\xi\, .
\end{equation}
Note that the damping kernel $D$ generally depends on $\bar\sigma$, too.

\section{The 2PI effective action for the linear sigma model with constituent quarks}\label{sec:2pieacal}

The formalism of the 2PI effective action \cite{Luttinger:1960ua,Lee:1960zza,Baym:1961zz,Baym:1962sx} is well-suited for our purpose because it yields a selfconsistent and thermodynamically consistent description of the entire system. Thus, it allows for a well defined back reaction of the relevant modes on the heat bath. In the scheme that is developed in the following we use the semiclassical approximation for the sigma field. We restrict ourselves to the sigma mean-field and do not include the propagator of the sigma field. We, thus, work with a theory of fermions coupled to an external mean field. Since the quarks have a vanishing mean field, they are represented by their propagators. Then the 2PI effective action is a functional of the sigma mean-field $\sigma^a(x)$ and the full quark propagator $S^{ab}(x,y)$
\begin{equation}
\Gamma[\sigma,S]=S_{\rm cl}[\sigma]-i{\rm Tr}\ln S^{-1}-i{\rm Tr}S_0^{-1}S+\Gamma_2[\sigma,S]\, ,
\label{eq:sc_gamma}
\end{equation}
where the trace operation includes ${\rm Tr}=\int_{\cal C}{\rm d}^4x\sum_{\rm flavor}\sum_{\rm Dirac}$ and $S_{\rm cl}[\sigma]$ is the classical action of the sigma mean-field. The free propagator for a fermion with mass $m_f$ inverts the differential operator of the free theory
\begin{equation}
 (i\slashed\partial-m_f)S_0^{ab}(x,y)=-i\delta_{\cal C}^{ab}(x-y)\, ,
\end{equation}
by which it is defined up to the boundary conditions.

The first three terms in (\ref{eq:sc_gamma}) are the one-loop results. The additional term $\Gamma_2[\sigma,S]$ is the sum of all 2PI diagrams.


In the absence of external sources the equation of motion for the sigma mean-field $\sigma^a$, obtained by variation of the effective action (\ref{eq:sc_gamma}) with respect to $\sigma^a$, is
\begin{equation}
 \frac{\delta\Gamma[\sigma,S]}{\delta\sigma^a}=0\, ,
\label{eq:sc_eomsigma}
\end{equation}
and for the full quark propagator $S^{ab}$, obtained by variation with respect to $S^{ab}$,
\begin{equation}
 \frac{\delta\Gamma[\sigma,S]}{\delta S^{ab}}=0\, .
\label{eq:sc_eomqprop}
\end{equation}
The proper self energy of the quarks is
\begin{equation}
 \Sigma^{ab}(x,y;S)=S_0^{ab}(x,y)^{-1}-S^{ab}(x,y)^{-1}\, .
\label{eq:sc_selfenergy1}
\end{equation}
Inserting the self energy (\ref{eq:sc_selfenergy1}) into the effective potential (\ref{eq:sc_gamma}) and neglecting constant terms gives
\begin{equation}
\Gamma[\sigma,S]=S_{\rm cl}[\sigma]-i{\rm Tr}\ln S^{-1}-i{\rm Tr}\Sigma S+\Gamma_2[\sigma,S]\, .
\label{eq:sc_gamma1}
\end{equation}
With $\delta\Sigma/\delta S=1/S^2$, the variation (\ref{eq:sc_eomqprop}) reads
\begin{equation}
 -i\Sigma^{ab}(x,y)=-\frac{\delta\Gamma_2[\sigma,S]}{\delta S^{ab}(x,y)}\, .
\label{eq:sc_selfenergy2}
\end{equation}
This means that the equation of motion for the full quark propagator $S^{ab}$ (\ref{eq:sc_eomqprop}) is equivalent to equation (\ref{eq:sc_selfenergy1}), where the self energy is given by the expression (\ref{eq:sc_selfenergy2}). From equation (\ref{eq:sc_selfenergy1}) we obtain the Schwinger-Dyson equation
\begin{equation}
 S_0^{-1}S-\Sigma S=1\, ,
\end{equation}
which is in explicit terms
\begin{equation}
 (i\slashed\partial -m_f)S^{ab}(x,y)-i\int_{\cal C}{\rm d}^4z\Sigma^{ac}(x,z)S^{cb}(z,y)=i\delta_{\cal C}^{ab}(x-y)\, .
\label{eq:sc_eomqprop2}
\end{equation}
 Since implicit dependencies are not varied, the equation of motion for the sigma mean-field is
\begin{equation}
 - \frac{\delta S_{\rm cl}[\sigma]}{\delta\sigma^a}=\frac{\delta\Gamma_2[\sigma,S]}{\delta\sigma^a}\, .
\label{eq:sc_eomsigma2}
\end{equation}
To solve the equation of motion for the quark propagator (\ref{eq:sc_eomqprop2}) and for the sigma mean-field (\ref{eq:sc_eomsigma2}) we need the explicit form of the self energy and, thus, with expression (\ref{eq:sc_selfenergy2}) the explicit form of $\Gamma_2[\sigma,S]$.

\subsection{The explicit form of $\Gamma_2[\sigma,S]$ and the self energy}
\begin{wrapfigure}{O}{0.3\textwidth}
  \begin{center}
   \includegraphics[bb=266 536 345 661]{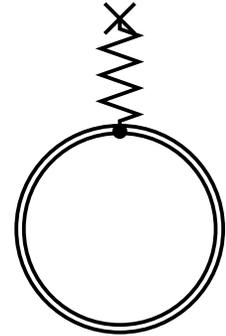}
  \end{center}
  \caption{The only diagram for $\Gamma_2[\sigma,S]$.}
 \label{fig:sc_gamma2} 
\end{wrapfigure}

Since all graphs with more than one mean-field insertion are necessarily two-particle reducible, they are not included in $\Gamma_2[\sigma,S]$. A single mean-field insertion represents them all, and we have to calculate only one diagram within the closed time path formalism, see figure \ref{fig:sc_gamma2}. 
For the inclusion of such types of diagrams in the $\Phi$-functional approach see \cite{Leupold:2006bp}. Note, that this is exact within the mean-field approximation for the sigma field, because there are no quark self-interactions in the theory that could contribute to the two-particle irreducible effective action. It consists of one graph with a $+$- and one with a $-$-vertex. The corresponding Feynman rules are taken from \cite{Ivanov:1998nv}.

The diagram is
\begin{equation}
 \Gamma_2[\sigma,S]=g\int_{\cal C}{\rm d}^4x\,{\rm tr}\,(S^{\scriptscriptstyle ++}(x,x)\sigma^+(x)+S^{\scriptscriptstyle --}(x,x)\sigma^-(x))\, ,
\label{eq:sc_gamma2exact}
\end{equation}
with the trace operation ${\rm tr}=\sum_{\rm flavor}\sum_{\rm Dirac}$. Then, the self energy from (\ref{eq:sc_selfenergy2}) reads
\begin{equation}
 \Sigma^{ab}(x,y)=-ig\delta_{\cal C}^{ab}(x-y)\sigma^b(x)\, .
\label{eq:sc_selfenergy3}
\end{equation}


\subsection{The coupled equations of motion}
With the explicit form of the self energy (\ref{eq:sc_selfenergy3}) the Schwinger-Dyson equation (\ref{eq:sc_eomqprop2}) 
\begin{equation}
 (i\slashed\partial -m_f)S^{ab}(x,y)-g\sigma^a(x)S^{ab}(x,y)=i\delta_{\cal C}^{ab}(x-y)\, ,
\label{eq:sc_eomqprop3}
\end{equation}
and the field equation for the sigma mean-field 
\begin{equation}
- \frac{\delta S_{\rm cl}[\sigma]}{\delta\sigma^a}=g\, {\rm tr}\, S^{aa}(x,x) 
\label{eq:sc_eomsigma3}
\end{equation}
are a coupled set of equations. In principle, we have to solve (\ref{eq:sc_eomqprop3}) and put the solution for the full propagator $S^{ab}$ into (\ref{eq:sc_eomsigma3}). Due to the space-time dependence of $\sigma^a$ in (\ref{eq:sc_eomqprop3}) it is generally nontrivial to find the solution for $S^{ab}$, which is exact for the given form of $\Gamma_2$ (\ref{eq:sc_gamma2exact}). We, therefore, have to approximate the full propagator. This is a crucial aspect because only the full approach of the two-particle irreducible effective action is a conserving, selfconsistent and thermodynamically consistent approximation to the exact quantum field theory \cite{Baym:1961zz,Baym:1962sx,Ivanov:1998nv}. For exact solutions of coupled propagator and mean-field dynamics for some model systems, see e.g. \cite{Berges:2001fi,Juchem:2003bi,Juchem:2004cs}.

We split the mean field into one component $\sigma^a_0$ that has a slow variation compared to $S^{ab}$ and a fluctuation part $\delta\sigma^a$, which we assume to be small. We will later disclose the actual meaning of this splitting
\begin{equation}
 \sigma^a(x)=\sigma^a_0(x)+\delta\sigma^a(x)\, .
\label{eq:sc_sigmasplitting}
\end{equation}
We also expand the full propagator around the thermal propagator
\begin{equation}
 S^{ab}(x,y)=S_{\rm th}^{ab}(x,y)+\delta S^{ab}(x,y)+\delta^2S^{ab}(x,y)\, .
\label{eq:sc_approxprop}
\end{equation}
Then for the various orders of the expansion the Schwinger-Dyson equation reads
\begin{subequations}
\begin{align}
 {\cal O}(0):\quad& (i\slashed\partial -m_f)S_{\rm th}^{ab}(x,y)-g\sigma^a_0(x)S_{\rm th}^{ab}(x,y)=i\delta_{\cal C}^{ab}(x-y)
\label{eq:sc_eomqpropo0}\\
 {\cal O}(1):\quad& (i\slashed\partial -m_f)\delta S^{ab}(x,y)-g\sigma^a_0(x)\delta S^{ab}(x,y)-g\delta\sigma^a(x) S_{\rm th}^{ab}(x,y)=0
\label{eq:sc_eomqpropo1}\\
 {\cal O}(2):\quad& (i\slashed\partial -m_f)\delta^2 S^{ab}(x,y)-g\sigma^a_0(x)\delta^2 S^{ab}(x,y)-g\delta\sigma^a(x) \delta S^{ab}(x,y)=0\, .
\label{eq:sc_eomqpropo2}
\end{align}
\end{subequations}
From (\ref{eq:sc_eomqpropo0}) we see that the $\sigma^a_0$ part of the sigma field generates the mass of the quarks dynamically $m=m_f+g\sigma_0$. As already noted the idea is that the $x$-dependence of $\sigma_0$ is weak compared to the $x$-dependence of the propagator S. We identify the spatial and temporal variation of sigma with the corresponding variation of the local temperature in the fluid dynamic description of the quarks and antiquarks. In that spirit the solution of equation  (\ref{eq:sc_eomqpropo0}) is given by 
\begin{subequations}
\label{eq:ftqft_fermiprop}
\begin{align}
iS^{++}(p)&=\langle T\psi(x)\psi^\dagger(x')\rangle=(\slashed p +m)\left(\frac{1}{p^2-m^2+i\epsilon}+2i\pi n_{\rm F}(|p^0|)\delta(p^2-m^2)\right)&\quad\text{for }t,t'\text{ on } {\cal C}_+\, ,
\label{eq:ftqft_fermiprop1}\\
iS^{+-}(p)&=\pm\langle \psi^\dagger(x')\psi(x)\rangle=2i\pi(\slashed p +m) (n_{\rm F}(|p^0|)-\Theta(-p^0))\delta(p^2-m^2)&\quad \text{for }t\text{ on } {\cal C}_+,t'\text{ on } {\cal C}_-\, ,
\label{eq:ftqft_fermiprop2}\\
iS^{-+}(p)&=\langle \psi(x)\psi^\dagger(x')\rangle=2i\pi(\slashed p +m) (n_{\rm F}(|p^0|)-\Theta(p^0))\delta(p^2-m^2)&\quad\text{for }t'\text{ on } {\cal C}_+,t\text{ on } {\cal C}_-\, ,
\label{eq:ftqft_fermiprop3}\\
iS^{--}(p)&=\langle T_a\psi(x)\psi^\dagger(x')\rangle=(\slashed p +m)\left(-\frac{1}{p^2-m^2-i\epsilon}+2i\pi n_{\rm F}(|p^0|)\delta(p^2-m^2)\right)&\quad\text{for }t,t'\text{ on } {\cal C}_-
\label{eq:ftqft_fermiprop4}\, .
\end{align}
\end{subequations}
From (\ref{eq:sc_eomqpropo1})
\begin{equation}
 \delta S^{ab}(x,y)=-ig\int_{\cal C}{\rm d}^4zS^{ac}_{\rm th}(x,z)\delta\sigma^c(z)S^{cb}_{\rm th}(z,y)\, ,
\label{eq:sc_delta1prop}
\end{equation}
and from (\ref{eq:sc_eomqpropo2})
\begin{equation}
 \delta^2 S^{ab}(x,y)=-g^2\int_{\cal C}{\rm d}^4z{\rm d}^4z'S^{ac}_{\rm th}(x,z')\delta\sigma^c(z')S^{cd}_{\rm th}(z',z)\delta\sigma^d(z)S^{db}_{\rm th}(z,y)\, .
\label{eq:sc_delta2prop}
\end{equation}
The approximated propagator (\ref{eq:sc_approxprop}) rewritten in center and relative variables
\begin{subequations}
 \begin{align}
 \delta\bar{\sigma}&=\frac{1}{2}(\delta\sigma^{\scriptscriptstyle +}+\delta\sigma^{\scriptscriptstyle -})\\
 \Delta\delta\sigma&=\delta\sigma^{\scriptscriptstyle +}-\delta\sigma^{\scriptscriptstyle -}
\end{align}
\end{subequations}
yields 
\begin{equation}
\begin{split}
  {\rm tr}S^{ab}(x,y)&={\rm tr}S_{\rm th}^{ab}(x,y)+ig\int_{y_0}^{x_0}\!\!{\rm d}^4y\delta\bar\sigma(y)\biggl(S^<(x-y)S^>(y-x)-S^>(x-y)S^<(y-x)\biggr)\\
		&\hphantom{={\rm tr}S_{\rm th}^{ab}(x,y)}-\frac{i}{2}g\int\!\!{\rm d}^4y\Delta\delta\bar\sigma(y)\biggl(S^<(x-y)S^>(y-x)+S^>(x-y)S^<(y-x)\biggr)\\
		&\stackrel{\mathclap{\Delta\delta\sigma=0}}{\rightarrow}\quad{\rm tr}S_{\rm th}^{ab}(x,y)+\frac{D(x)}{g}\, .
\end{split}
\label{eq:sc_approxpropexplicit}
\end{equation}
In the last step we identified the same damping kernel as in (\ref{eq:sc_dampingkernel}). A term similar to the noise kernel in (\ref{eq:sc_noisekernel}) vanished by taking $\Delta\delta\sigma=0$. In order to recover the noise kernel we need to calculate the effective action $\Gamma[\sigma,S]$ explicitly from  (\ref{eq:sc_gamma}) with the approximations of the propagator (\ref{eq:sc_approxprop}) and (\ref{eq:sc_delta1prop}-\ref{eq:sc_delta2prop}).
It is
\begin{equation}
\begin{split}
  \Gamma[\sigma,S]&=S_{\rm cl}[\sigma]+g{\rm tr}\;S^{\scriptscriptstyle ++}_{\rm th}(x,x)\Delta\sigma(x)\\
		&\hphantom{=}+\frac{i}{2}g^2\int\!\!{\rm d}^4x\int_{y_0}^{x_0}\!\!{\rm d}^4y\Delta\delta\sigma(x)\delta\bar\sigma(y)\biggl(S^<(x-y)S^>(y-x)-S^>(x-y)S^<(y-x)\biggr)\\
    &\hphantom{=}-\frac{i}{4}g^2\int\!\!{\rm d}^4x\int\!\!{\rm d}^4y\Delta\delta\sigma(x)\Delta\delta\sigma(y)\biggl(S^<(x-y)S^>(y-x)+S^>(x-y)S^<(y-x)\biggr)
\end{split}
\label{eq:sc_gammaexp}
\end{equation}

We can readily identify the same damping (\ref{eq:sc_dampingkernel}) and noise kernel (\ref{eq:sc_noisekernel}) as we found in the influence functional approach.

The equation of motion for the sigma mean-field obtained by varying $\Gamma[\sigma,S]$ with respect to $\Delta\sigma$  is
\begin{equation}
 -\frac{\delta S_{\rm cl}[\sigma,S]}{\delta\sigma^a}=g{\rm tr}\;S^{++}_{\rm th}(x,x) +D(x)+\xi(x)\, ,
\label{eq:sc_eomsigma5}
\end{equation}
where we introduced the same stochastic field as was discussed in section \ref{sec:noisekernel}. Note, that within the 2PI effective action we obtain the standard mean-field result as the first term on the right hand side of equation (\ref{eq:sc_eomsigma5}), which is formally of order $g$, while the damping and the noise kernel are of order $g^2$. Therefore, the standard mean-field result is the lowest order contribution.

\subsection{The thermodynamic quantities of the quark fluid}
Using the 2PI effective action we can include the local equilibrium properties of the quark fluid. In (\ref{eq:sc_gammaexp}) the calculations were performed along the real-time contour from figure \ref{fig:qft_contour}. For the calculation of equilibrium properties of a thermodynamic system we cannot, however, neglect the imaginary-time path ${\cal C}_3$. 
 In the imaginary-time formalism the thermodynamic potential can be evaluated in a diagrammatic expansion \cite{Kap94}. The perturbative expansion of the thermodynamic potential in real-time formalism is more difficult. The Bogoliubov assumption of an uncorrelated initial state  \cite{Bog62} leads to the factorization of the generating functional \cite{Das97}. For the derivation of the Green's functions and the dynamics of the system the imaginary-time path ${\cal C}_3$ of the contour can be neglected. To obtain the correct equilibrium properties of the coupled system it must be included since it contributes to the pressure
\begin{equation}
 p=\frac{T}{V}\left(\ln Z_{\cal C}+\ln Z_3\right)=-\Gamma[\sigma,S]+\frac{T}{V}\ln Z_3\, .
\end{equation}
In equilibrium with $\Gamma[\sigma,S]|_{\Delta\sigma=0}=0$ the full  pressure is given by the imaginary-time path ${\cal C}_3$ of the full contour, see figure \ref{fig:qft_contour}. Due to the Bogoliubov initial conditions we are left only with the one-loop effective potential. More advanced techniques are required to set up a consistent real-time perturbation expansion for equilibrium properties \cite{Landsman:1986uw}. Since we restricted the model to the mean-field dynamics higher loop corrections to the pressure associated with propagators of the sigma field are discarded in the entire setting. This issue was briefly mentioned in the beginning of this section. Here, it assures that we capture the full equilibrium properties by the mean-field pressure calculated in equation (\ref{eq:pressure}).

\begin{figure}
  \centering
 \includegraphics[bb=107 448 461 609]{CTP.eps}
\caption{The complete time path contour with the real time paths ${\cal C}_+$ and ${\cal C}_-$ and the imaginary time path ${\cal C}_3$.}
\label{fig:qft_contour}
\end{figure}

\section{The equation of motion for the sigma field}\label{sec:sc_eomsigmafield}
We now turn to the explicit calculation of the terms in the equation of motion for the sigma field (\ref{eq:sc_eomsigma5}). On the right hand side, it includes the lowest order contribution, a damping term and the correlation of the noise.

The free fermionic propagators on the real-time contour are given by equation (\ref{eq:ftqft_fermiprop}).

In coordinate space the propagators obey the relations
\begin{subequations}
\begin{align}
 S^{++}(x-y)&=S^{-+}(x-y)\Theta(x_0-y_0)+S^{+-}(x-y)\Theta(y_0-x_0)\label{eq:ftqft_proprel1} \\
 S^{--}(x-y)&=S^{+-}(x-y)\Theta(x_0-y_0)+S^{-+}(x-y)\Theta(y_0-x_0)\label{eq:ftqft_proprel2}\, .
\end{align}
\end{subequations}

\subsection{Lowest order}
For the lowest order contribution we calculate the first term on the right hand side of equation (\ref{eq:sc_eomsigma5}) with the thermal part of the free quark propagator (\ref{eq:ftqft_fermiprop1})
\begin{equation}
\begin{split}
 g\,{\rm tr}S^{\scriptscriptstyle ++}(x,x)&=ig\int\frac{{\rm d}^4p}{(2\pi)^4}{\rm tr}\;2i\pi n_{\rm F}(|p^0|)(\gamma^\mu p_\mu+m_q(x))\delta(p^2-m_q^2)\\
	  &=-2d_qg^2\sigma(x)\int\frac{{\rm d}^3p}{(2\pi)^3}\frac{n_{\rm F}(E_p)}{E_p}\\
          &=-g\rho_s(x)
\end{split}
\label{eq:sc_lowestorder}
\end{equation}
with the dynamically generated quark mass $m_q(x)=g\sigma(x)$ and the degeneracy factor $d_q=12$ from the trace over flavor, color and spin. The energy of the quarks is $E_p=\sqrt{\vec p^2+m_q^2}$ and $\rho_s$ is the one-loop scalar density. We see that to this order the equation of motion does not include any terms leading to damping and noise. It is the same classical equation of motion that was used in previous chiral fluid dynamic models.

\subsection{The damping kernel}\label{sec:dampingterm}
The explicit form of the damping kernel $D(x)$ is given in equation (\ref{eq:sc_dampingkernel}). For its evaluation we define the following quantity
\begin{equation}
 {\cal M}(x-y)={\rm tr} \left(S^{<}(x-y)S^{>}(y-x)-S^{>}(x-y)S^{<}(y-x)\right)\, .
\end{equation}
Its Fourier transform can be evaluated, tracing over flavor, color and spin, and performing the integration over $p^0$ by the use of the delta functions from (\ref{eq:ftqft_fermiprop2}) and (\ref{eq:ftqft_fermiprop3}). 

%
We sort the various scattering processes according to their energy balance, given by the delta functions, in order to make the physical processes obvious. We obtain
\begin{equation}\begin{split}
 {\cal M}(\omega,\bf k)=&-\frac{d_q}{4\pi^2}\int{\rm d}^3p\frac{1}{E_pE_{k+p}}{\scriptstyle \times}\\
&{\scriptstyle \times} \biggl(\!(-(E_p+E_{k+p})E_p+2m_q^2-{\bf k}\cdot{\bf p}){\scriptstyle \times} \\
  &\hphantom{{\scriptstyle \times}\biggl(}\vphantom{\biggl(}{\scriptstyle \times}\left\{\delta(\omega+E_p+E_{k+p})(n_{\rm F}(E_{k+p})n_{\rm F}(E_p)-(1-n_{\rm F}(E_p))(1-n_{\rm F}(E_{k+p})))\right.\\
 &\hphantom{{\scriptstyle \times\biggl({\scriptstyle \times}(}}\vphantom{\biggl(}\left.+\delta(\omega-E_p-E_{k+p})((1-n_{\rm F}(E_p))(1-n_{\rm F}(E_{k+p}))-n_{\rm F}(E_{k+p})n_{\rm F}(E_p))\right\}\\
   &\hphantom{{\scriptstyle \times}(}\vphantom{\biggl(}+((-E_p+E_{k+p})E_p+2m_q^2-{\bf k}\cdot{\bf p}){\scriptstyle \times}\\
 &\hphantom{{\scriptstyle \times}\biggl(}\vphantom{\biggl(}{\scriptstyle \times}\left\{\delta(\omega+E_p-E_{k+p})((1-n_{\rm F}(E_p))n_{\rm F}(E_{k+p})-n_{\rm F}(E_p)(1-n_{\rm F}(E_{k+p})))\right.\\
&\hphantom{{\scriptstyle \times}\biggl({\scriptstyle \times}(}\left.+\delta(\omega-E_p+E_{k+p})(n_{\rm F}(E_p)(1-n_{\rm F}(E_{k+p}))-(1-n_{\rm F}(E_p))n_{\rm F}(E_{k+p}))\right\}\!\biggr)
\end{split}
\label{eq:sc_M3}
\end{equation}
Here, one sees the antisymmetric property ${\cal M}(-\omega,{\bf k})=-{\cal M}(\omega,{\bf k})$. The structure of ${\cal M}(\omega,\bf k)$ is the following. It describes the difference between a gain and a loss term. For example, the term $n_{\rm F}(E_{k+p})n_{\rm F}(E_p)$ is the probability for a quark-antiquark pair to form a sigma mode $\bar qq\to\sigma$, and the term $(1-n_{\rm F}(E_p))(1-n_{\rm F}(E_{k+p}))$ is the statistical weight of the decay of a sigma mode to a quark-antiquark pair $\sigma\to\bar qq$. The mixed terms $(1-n_{\rm F}(E_p))n_{\rm F}(E_{k+p})$ describe the scattering of a quark (antiquark) off a sigma mode to form an antiquark (quark). For each delta function, the ratio of the loss to the gain term is
\begin{equation}
 \frac{\Gamma_{\rm loss}}{\Gamma_{\rm gain}}=\exp\left(\frac{\omega}{T}\right)\, .
\end{equation}
This is the detailed balance relation for the thermal quarks and antiquarks. For fixed quark masses the delta functions constrain the allowed scattering processes \cite{Weldon:1983jn}. 

The damping kernel is then
\begin{equation}
\begin{split}
D(x)&=ig^2\int_{y_0}^{x_0}{\rm d}^4y{\cal M}(x-y)\delta\bar\sigma(y)\\
    &=ig^2\int_{y_0}^{x_0}{\rm d}^4y\int\frac{{\rm d}^4k}{(2\pi)^4}\exp(-ik(x-y)){\cal M}(\omega,{\bf k})\delta\bar\sigma(y)\, , 
\end{split}
\end{equation}
where we take the spatial Fourier transform
\begin{equation}
 \delta\bar\sigma(y_0,{\bf k})=\int{\rm d}^3y\exp(-i{\bf k}\cdot{\bf y})\delta\bar\sigma(y_0,{\bf y})\, .
\end{equation}
To simplify further calculations we make the substitution $y_0=x_0-\tau$ and assume that the initial time $x_0-y_0\to-\infty$, such that $\tau\to\infty$. Then
\begin{equation}
 D(x)=ig^2\int\frac{{\rm d}^3k}{(2\pi)^3}\exp(i{\bf k}\cdot{\bf x})\int_0^\infty{\rm d}\tau\int\frac{{\rm d}\omega}{(2\pi)}\exp(-i\omega\tau){\cal M}(\omega,{\bf k})\delta\bar\sigma(x_0-\tau,{\bf k})\, .
\end{equation}
We see that the sigma mean-field now depends on the history $t<x_0$. It is known that an instantaneous approximation to this time dependence is too constraining because the dissipative terms vanish \cite{Morikawa:1986rp,Gleiser:1993ea}. We assume that the effect of the past can be described by harmonic oscillations around a constant value
\begin{equation}
 \bar\sigma(t-\tau,{\bf k})\simeq a(t)\cos(E_k\tau)+b(t)\sin(E_k\tau)\, .
\label{eq:sc_linharmapp1}
\end{equation}
This is the linear harmonic approximation which is also used in \cite{Greiner:1996dx,Rischke:1998qy}. Here, we obtain the coefficients from the requirements at $\tau=0$
\begin{alignat}{2}
 \bar\sigma(t-\tau,{\bf k})|_{\tau=0}&=\bar\sigma_0(t,{\bf k})\quad&\Rightarrow&\quad a(t)=\bar\sigma_0(t,{\bf k})\,\\
\frac{\partial\bar\sigma(t-\tau,{\bf k})}{\partial \tau}|_{\tau=0}&=-\frac{\partial\bar\sigma(t,{\bf k})}{\partial t}\quad&\Rightarrow&\quad b(t)=-\frac{1}{E_k}\partial_t\bar\sigma(t,{\bf k})\, .
\end{alignat}
Then, we see that
\begin{equation}
\begin{split}
  \bar\sigma(t-\tau,{\bf k})&\simeq \sigma_0(t,{\bf k}) \cos(E_k\tau)-\frac{1}{E_k}\partial_t\bar\sigma(t,{\bf k})\sin(E_k\tau)\\
			&=\sigma_0(t,{\bf k})+\delta\bar\sigma(t,{\bf k}) \, .
\end{split}
\end{equation}
We now see the meaning of the splitting of $\sigma(x)$ that was done in (\ref{eq:sc_sigmasplitting}) and
\begin{equation}
 \delta\bar\sigma(t,{\bf k})=\sigma_0(t,{\bf k})(\cos(E_k\tau)-1)-\frac{1}{E_k}\partial_t\bar\sigma(t,{\bf k})\sin(E_k\tau)\, .
\label{eq:sc_linharmapp2}
\end{equation}
The first term gives a mass shift for the sigma field, which is only a correction to the leading order result (\ref{eq:sc_lowestorder}). We assume that this correction is small as $\cos(E_k\tau)-1\simeq0$. Then, we can replace the fluctuation $\delta\bar\sigma(t,{\bf k})$ by the sine-modulated time derivative of the full field.
 With this we can evaluate the integral over the history and obtain quantities that are local in time. Such an approximation will later be used for the derivation of the noise correlator, too. 
Writing
\begin{equation}
 \delta\bar\sigma(t-\tau,{\bf k})=-\frac{1}{2iE_k}(\exp(iE_k\tau)-\exp(-iE_k\tau))\partial_t\bar\sigma(t,{\bf k})
\end{equation}
and using the relation
\begin{equation}
 \int_0^\infty{\rm d}\tau \exp(i(\omega-E)\tau)=i{\cal P}\frac{1}{\omega-E}+\pi\delta(\omega-E)
\end{equation}
we arrive at
\begin{equation}
 \begin{split}
  D(x)&=-g^2\int\frac{{\rm d}^3k}{(2\pi)^3}\exp(i{\bf k}\cdot{\bf x})\int\frac{{\rm d}\omega}{(2\pi)}{\cal M}(\omega,{\bf k}){\scriptstyle\times}\\
 &\hphantom{=-g^2\int}{\scriptstyle\times}\frac{1}{2E_k}\left(i{\cal P}\frac{1}{E_k-\omega}+\pi\delta(E_k-\omega)-i{\cal P}\frac{1}{-\omega-E_k}-\pi\delta(-\omega-E_k))\partial_t\bar\sigma(t,{\bf k})\right)\\
&=-g^2\int\frac{{\rm d}^3k}{(2\pi)^3}\exp(i{\bf k}\cdot{\bf x})\int\frac{{\rm d}\omega}{(2\pi)}{\cal M}(\omega,{\bf k})\frac{\pi}{E_k}\delta(\omega-E_k)\partial_t\bar\sigma(t,{\bf k})\, .
 \end{split}
\end{equation}
In the final step, we used that the principle integral terms cancel by applying the antisymmetry of ${\cal M}(\omega,{\bf k})$. We obtain for the damping kernel
\begin{equation}
 D(x)=-g^2\int\frac{{\rm d}^3 k}{(2\pi)^3}\exp(i{\bf k}\cdot{\bf x})\frac{1}{2E_k}{\cal M}(E_k,{\bf k})\partial_t \bar\sigma(t,{\bf k})\, .
\label{eq:sc_dampkern}
\end{equation}

${\cal M}(\omega,{\bf k})$ contains the on-shell reaction rate of the processes given in equation (\ref{eq:sc_M3}). They lead to the dissipative part of the equation of motion. 

In a perturbative expansion the damping term appears first at next-to-leading order $g^2$ as one can immediately read off from (\ref{eq:sc_dampkern}). Also the thermal mass correction, which we have neglected is of this order $g^2$. However, it is in fact only a correction to the mass of the sigma meson, which gets contributions from leading order (\ref{eq:sc_lowestorder}), i.e. from the standard mean-field contribution, and even more from the sigma field potential (\ref{eq:Uchi}). 



\begin{figure}
  \centering
  \includegraphics{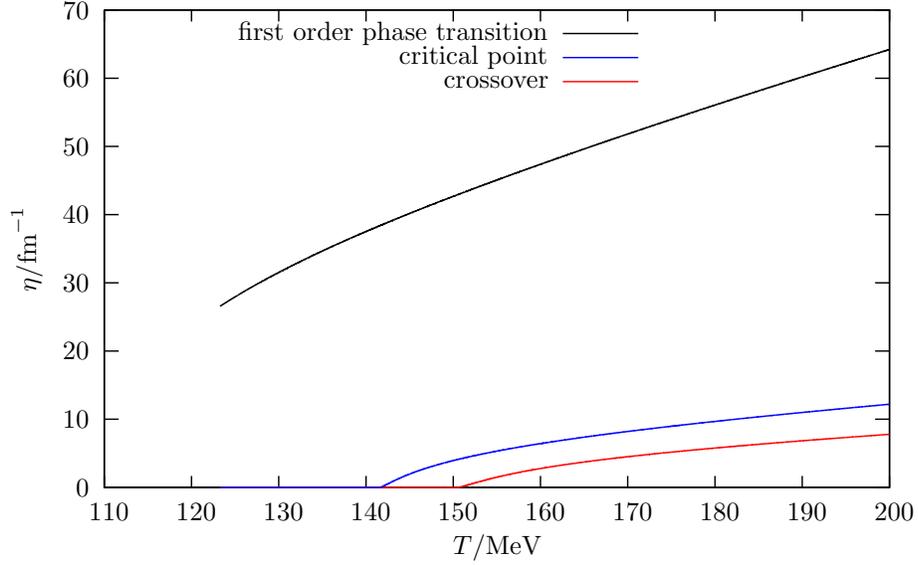}
  \caption{Temperature dependence of the damping coefficient $\eta$ for a the different couplings $g=5.5$, $g=3.63$ and $g=3.3$, which correspond to scenarios with a first order phase transition, a critical point and a crossover.}
  \label{fig:sc_dampfung}
 \end{figure}

Being interested in the long-range oscillations of sigma we calculate the damping coefficient $\eta$ for the zero mode, ${\bf k}=0$, of the sigma mean-field and approximate ${\cal M}(E_k,{\bf k})\simeq{\cal M}(m_\sigma,0)$. Then for $m_\sigma>2m_q$ only the process $\sigma\rightarrow\bar q q$ and the reverse reaction $\bar q q\rightarrow\sigma$ are kinematically possible. We find
\begin{equation}
 \begin{split}
{\cal M}(m_\sigma,0)&=-\frac{d_q}{2\pi^2}\int{\rm d}^3p\frac{(m^2-E_p^2)}{E_p^2}(1-2n_{\rm F}(E_p))\delta(m_\sigma-2E_p)\\
 		    &=2\frac{d_q}{\pi}\left(1-2n_{\rm F}\left(\frac{m_\sigma}{2}\right)\right)\frac{1}{m_\sigma}\left(\frac{m_\sigma^2}{4}-m_q^2\right)^{3/2}\, .
 \end{split}
\label{eq:sc_Minapprox}
\end{equation}
With the same approximation $E_k\simeq m_\sigma$ the damping kernel becomes
 \begin{equation}
  \begin{split}
  D(x)&\simeq-g^2\int\frac{{\rm d}^3 k}{(2\pi)^3}\exp(i{\bf k}\cdot{\bf x})\frac{1}{2m_\sigma}{\cal M}(m_\sigma,0)\partial_t \bar\sigma(t,\bf k)\\
&=-g^2\frac{d_q}{\pi}\left(1-2n_{\rm F}\left(\frac{m_\sigma}{2}\right)\right)\frac{1}{m_\sigma}\left(\frac{m_\sigma^2}{4}-m_q^2\right)^{3/2}\partial_t \bar\sigma(t,{\bf x})\, .
 \end{split}
\label{eq:sc_dampingkernelapp}
 \end{equation}
With the equation of motion (\ref{eq:sc_eomsigma5}) the damping coefficient can be identified as
\begin{equation}
\eta= g^2\frac{d_q}{\pi}\left(1-2n_{\rm F}\left(\frac{m_\sigma}{2}\right)\right)\frac{1}{m_\sigma}\left(\frac{m_\sigma^2}{4}-m_q^2\right)^{3/2}\, .
\label{eq:sc_dampingcoefficient}
\end{equation}
Its temperature dependence is shown in figure \ref{fig:sc_dampfung} for the three different phase transition scenarios. The value of the sigma field in $m_q=g\sigma$ and the sigma mass $m_\sigma$ are taken to be the equilibrium values at the given temperature. Since $\eta\propto g^2$ it is larger in a first order scenario than in a scenario with a critical point. This issue appears because we work at $\mu=0$ and tune the strength of the phase transition by different values of the coupling $g$. In the linear sigma model with constituent quarks a realistic constituent quark mass is obtained for $g=3.3$.  
Though the linear sigma model does not include confinement the damping coefficient for the zero mode of the sigma field obtained from the interaction with the quarks vanishes below the phase transition. This gives a realistic description at low temperatures.
The reason is that at high temperatures the (mostly dynamically generated) quark mass is small and therefore $m_\sigma > 2 m_q$ is satisfied. Hence the reactions $\sigma \leftrightarrow \bar q q$ can take place. With lower temperatures $m_q$ rises and at some point the reactions, which cause damping and noise in our model are kinematically forbidden. Physically, we expect that at low temperatures the decay and formation processes $\sigma \leftrightarrow 2 \pi$ become important since the pions as quasi-Goldstone bosons of chiral symmetry breaking become very light. In the present approach we have neglected the pions. This will be improved in the future. In that context it is interesting to note that our values for the damping are very large compared to the ones deduced from the linear sigma model {\em without} quarks \cite{Biro:1997va,Rischke:1998qy}.
%


\subsection{Correlation of the noise fields}\label{sec:noiseterm}

For the derivation of the correlation of the noise fields we perform the same steps for the noise kernel (\ref{eq:sc_noisekernel}) as for ${\cal M}(x-y)$.
For the Fourier transform of (\ref{eq:sc_noisekernel}) we find analogously
\begin{equation}
 \begin{split}
 {\cal N}(\omega,{\bf k})&=\frac{d_q}{4\pi^2}\int{\rm d}^3p\frac{1}{E_pE_{k+p}}{\scriptstyle \times}\\
&{\scriptstyle \times} \biggl(\!(-(E_p+E_{k+p})E_p+2m_q^2-{\bf k}\cdot{\bf p}){\scriptstyle \times} \\
  &\hphantom{{\scriptstyle \times}\biggl(}\vphantom{\biggl(}{\scriptstyle \times}\left\{\delta(\omega+E_p+E_{k+p})(n_{\rm F}(E_{k+p})n_{\rm F}(E_p)+(1-n_{\rm F}(E_p))(1-n_{\rm F}(E_{k+p})))\right.\\
 &\hphantom{{\scriptstyle \times\biggl({\scriptstyle \times}(}}\vphantom{\biggl(}\left.+\delta(\omega-E_p-E_{k+p})((1-n_{\rm F}(E_p))(1-n_{\rm F}(E_{k+p}))+n_{\rm F}(E_{k+p})n_{\rm F}(E_p))\right\}\\
   &\hphantom{{\scriptstyle \times}(}\vphantom{\biggl(}+((-E_p+E_{k+p})E_p+2m_q^2-{\bf k}\cdot{\bf p}){\scriptstyle \times}\\
 &\hphantom{{\scriptstyle \times}\biggl(}\vphantom{\biggl(}{\scriptstyle \times}\left\{\delta(\omega+E_p-E_{k+p})((1-n_{\rm F}(E_p))n_{\rm F}(E_{k+p})+n_{\rm F}(E_p)(1-n_{\rm F}(E_{k+p})))\right.\\
&\hphantom{{\scriptstyle \times}\biggl({\scriptstyle \times}(}\left.+\delta(\omega-E_p+E_{k+p})(n_{\rm F}(E_p)(1-n_{\rm F}(E_{k+p}))+(1-n_{\rm F}(E_p))n_{\rm F}(E_{k+p}))\right\}\!\biggr)
\end{split}\, .
\end{equation}
Since the noise term has the same microscopic origin as the damping term it is not surprising that the structure is very similar to (\ref{eq:sc_M3}). Especially, we find that ${\cal N}(\omega,{\bf k})$ is proportional to the sum of the loss and the gain term of the same scattering processes. 
The variance of the noise fields is
 \begin{equation}
  \begin{split}
  \langle\xi(t,{\bf x})\xi(t',{\bf x}')\rangle_\xi&= {\cal N}(x,y)\\
 						&=\int\frac{{\rm d}^4 k}{(2\pi)^4}{\cal N}(\omega,{\bf k})\exp(-i\omega(t-t'))\exp(i{\bf k}\cdot({\bf x}-{\bf x}'))\, ,
  \end{split}
 \end{equation}
where the average $\langle\rangle_\xi$ is taken with respect to the Gauss distribution (\ref{eq:sc_gaussdissxi}). With the approximation ${\cal N}(\omega,{\bf k})\simeq{\cal N}(m_\sigma,0)$ the integral over ${\bf k}$ can be evaluated to yield a spatial delta-function, which is replaced by the inverse volume term.
Then
 \begin{equation}
  \begin{split}
  \langle\xi(t)\xi(t')\rangle_\xi&=\frac{1}{V}\int\frac{{\rm d}\omega}{2\pi}{\cal N}(m_\sigma,0)\exp(-i\omega(t-t'))\\
 				&=\frac{1}{V}{\cal N}(m_\sigma,0)\delta(t-t')\, .
  \end{split}
\label{sc_noisecorrelator1}
 \end{equation}
With
\begin{equation}
 2n_{\rm F}^2(E_p)-2n_{\rm F}(E_p)+1=(1-2n_{\rm F}(E_p))\coth\left(\frac{E_p}{T}\right)
\end{equation}
the relation between ${\cal N}(m_\sigma,0)$ and ${\cal M}(m_\sigma,0)$ is
\begin{equation}
 {\cal N}(m_\sigma,0)={\cal M}(m_\sigma,0){\rm coth}\left(\frac{m_\sigma}{2T}\right)\, .
\end{equation}

With the explicit form of ${\cal M}(m_\sigma,0)$ (\ref{eq:sc_Minapprox}) and the damping coefficient (\ref{eq:sc_dampingcoefficient}), we finally find
\begin{equation}
 \langle\xi(t)\xi(t')\rangle_\xi=\frac{1}{V}\delta(t-t')m_\sigma\eta\coth\left(\frac{m_\sigma}{2T}\right)\, .
\label{eq:sc_noisecorrelation}
\end{equation}
The approximation ${\bf k}=0$ and, thus, $\omega=m_\sigma$ leads to the delta-function in the noise correlator. The noise fields are only correlated for equal times. This is the white-noise or Markovian approximation. In the non-Markovian description ${\cal N}(\omega, {\bf k})$ and ${\cal M}(\omega, {\bf k})$ have a full dependence on $k$. The delta-function in (\ref{sc_noisecorrelator1}) is replaced by a noise kernel that includes the memory effects of the history of the noise fields. In addition, the damping kernel in (\ref{eq:sc_dampingkernelapp}) is nonlocal. Non-Markovian noises pose a difficult problem for numerical studies \cite{Xu:1999aq}. 

\parskip 10pt
Finally, the equation of motion for the sigma field is
\begin{equation}
  \partial_\mu\partial^\mu\sigma+\frac{\delta U}{\delta\sigma}+g\rho_s+\eta\partial_t \bar\sigma(x)=\xi(x)
\end{equation}
with the scalar density $\rho_s$ defined in from (\ref{eq:sc_lowestorder}), the damping coefficient $\eta$ given in (\ref{eq:sc_dampingcoefficient}) and the correlation of the noise field $\xi$ given in (\ref{eq:sc_noisecorrelation}).
\parskip 0pt

\section{Energy-momentum conservation}\label{sec:sc_energyconservation}
From the Lagrangian (\ref{eq:LGML}) we calculate the divergence of the total averaged energy momentum tensor. On the operator level we have the Dirac equation for the quark operator and the conjugate for the adjoint operator
\begin{equation}
 (i\slashed\partial -g\sigma)q=0\quad{\rm and}\quad \bar q(i\overset{\leftarrow}{\slashed\partial}+g\sigma)=0\, .
\label{eq:sc_diraceq}
\end{equation}
Then, the energy-momentum tensor for the quarks reads
\begin{equation}
 T_q^{\mu\nu}(x)=i\bar q(x)\gamma^\mu\partial^\nu q(x)\, .
\label{eq:sc_tmunu}
\end{equation}
Taking the divergence yields
\begin{equation}
\begin{split}
 \partial_\mu T_q^{\mu\nu}&=i\bar q\overset{\leftarrow}{\slashed\partial}\partial^\nu q+i\bar q\partial^\nu\overset{\rightarrow}{\slashed\partial} q\\
				&=g\bar qq\partial^\nu\sigma\, ,
\end{split}
\label{eq:sc_tmunuquarkdiv}
\end{equation}
where we used the Dirac equation (\ref{eq:sc_diraceq}). The energy-momentum tensor for the sigma field can easily be derived from the purely mesonic part of the Lagrangian (\ref{eq:LGML})
\begin{equation}
 {\cal L}_\sigma=\frac{1}{2}\partial_\mu\sigma\partial^\mu\sigma - U(\sigma,{\vec \pi}=0)\, .
\end{equation}
The equation of motion for the sigma field is then found by the variational principle
\begin{equation}
 \partial_\mu\partial^\mu\sigma+\frac{\delta U}{\delta\sigma}+g\bar qq=0\, .
\end{equation}
The divergence of the energy-momentum tensor for the sigma field is
\begin{equation}
 \partial_\mu T_\sigma^{\mu\nu}=\left(\partial_\mu\partial^\mu\sigma+\frac{\delta U}{\delta\sigma}\right)\partial^\nu\sigma=-g\bar qq\partial^\nu \sigma\, .
\label{eq:sc_tmunusigdiv}
\end{equation}
From (\ref{eq:sc_tmunuquarkdiv}) and (\ref{eq:sc_tmunusigdiv}) it is clear that the sum $\partial_\mu T_q^{\mu\nu}+\partial_\mu T_\sigma^{\mu\nu}=0$ and, thus, the total energy of the system is conserved. 
Within the full formalism of the two-particle irreducible effective action we can now take the ensemble averages of the calculated quantities. Since the sigma field is treated in mean-field approximation we find with (\ref{eq:ftqft_fermiprop1}),  
\begin{equation}
 \partial_\mu T_q^{\mu\nu}(x)=-S^{\scriptscriptstyle ++}(x,x)\partial^\nu_x\sigma(x)
\label{eq:sc_tmunuquarkdiv2}
\end{equation}
and
\begin{equation}
 \partial_\mu T_\sigma^{\mu\nu}=S^{\scriptscriptstyle ++}(x,x)\partial^\nu_x\sigma(x)\, .
\label{eq:sc_tmunusigdiv2}
\end{equation}
The total averaged energy-momentum tensor is, thus, conserved,
\begin{equation}
 T_{\rm total}^{\mu\nu}=T_q^{\mu\nu}(x)+T_\sigma^{\mu\nu}(x)\, .
\label{eq:sc_totaltmunu}
\end{equation}

 The situation is more difficult for any approximation to the full propagator due to the space-time dependence of the effective mass generated by the dynamic symmetry breaking. We write the energy-momentum tensor of the quarks in the symmetric form
\begin{equation}
  \begin{split}
T_q^{\mu\nu}(x)&=\frac{i}{4}\bar q(x)(\gamma^\mu\partial^\nu q(x)+\gamma^\nu\partial^\mu q(x))-\frac{i}{4}(\partial^\mu\bar q(x)\gamma^\nu+\partial^\nu\bar q(x)\gamma^\mu)q(x)\\
  	&=\frac{i}{4}\left(\partial^\nu_x\bar q(y)q(x)|_{y=x}\gamma^\mu+\partial^\mu_x\bar q(y)q(x)|_{y=x}\gamma^\nu\right.\\
	&\hphantom{=\frac{i}{4}(}\left.-\partial^\mu_y\bar q(y)q(x)|_{x=y}\gamma^\nu-\partial^\nu_y\bar q(y)q(x)|_{x=y}\gamma^\mu\right)\, .
  \end{split}
\end{equation}
After a transformation to center and relative variable $X=1/2(x+y)$ and $u=x-y$ we see that the differentiation with respect to the center variable cancels and the remaining expression for the energy-momentum tensor of the quarks reads
\begin{equation}
 T_q^{\mu\nu}(X)=-\frac{i}{2}\partial^\nu_u S^{\scriptscriptstyle +-}(X,u)|_{u=0}\gamma^\mu-\frac{i}{2}\partial^\mu_u S^{\scriptscriptstyle +-}(X,u)|_{u=0}\gamma^\nu\, .
\end{equation}
The energy-momentum tensor of the sigma field remains the same by using the equation of motion (\ref{eq:sc_eomsigma3}) defined on the center variable
 \begin{equation}
 \partial_\mu T_\sigma^{\mu\nu}(X)=\left(\partial_\mu\partial^\mu\sigma+\frac{\delta U}{\delta\sigma}\right)\partial^\nu\sigma=\frac{\delta\Gamma_2}{\delta\sigma}\partial^\nu\sigma=S^{\scriptscriptstyle ++}(X)\partial^\nu_X\sigma(X)\, .
\end{equation}

With the approximation to the full propagator (\ref{eq:sc_approxprop}) to first order (\ref{eq:sc_delta1prop}) the energy-momentum balance of the entire system reads
\begin{equation}
\begin{split}
 \partial_\mu T_{q,{\rm appr.}}^{\mu\nu}(X)+\partial_\mu T_{\sigma, {\rm appr.}}^{\mu\nu}(X)&=\partial_\mu(-\frac{i}{2}\partial^\nu_u S_{\rm th}^{\scriptscriptstyle +-}(X,u)|_{u=0}\gamma^\mu-\frac{i}{2}\partial^\mu_u S_{\rm th}^{\scriptscriptstyle +-}(X,u)|_{u=0}\gamma^\nu\\
 	&\hphantom{=\partial_\mu(}-\frac{i}{2}\partial^\nu_u \delta S^{\scriptscriptstyle +-}(X,u)|_{u=0}\gamma^\mu-\frac{i}{2}\partial^\mu_u \delta S^{\scriptscriptstyle +-}(X,u)|_{u=0}\gamma^\nu)\\
 	&\hphantom{=}+S_{\rm th}^{\scriptscriptstyle ++}(X)\partial^\nu_X\sigma(X)+\delta S^{\scriptscriptstyle ++}(X)\partial^\nu_X\sigma(X)\, ,
 \end{split}
\label{eq:sc_approxTmunu}
\end{equation}
where we identify the four contributions
\begin{equation}
 \partial_\mu T_{q,{\rm appr.}}^{\mu\nu}(X)+\partial_\mu T_{\sigma, {\rm appr.}}^{\mu\nu}(X)=\partial_\mu T_{q,{\rm th}}^{\mu\nu}(X) +\partial_\mu T_{\sigma, {\rm th}}^{\mu\nu}(x)+\partial_\mu\delta T_q^{\mu\nu}(X)+\partial_\mu \delta T_{\sigma}^{\mu\nu}(x)
\end{equation}
The first term evaluates to
\begin{equation}
\begin{split}
 T_{q,{\rm th}}^{\mu\nu}(X)&=-\frac{i}{2}\partial^\nu_u S_{\rm th}^{\scriptscriptstyle +-}(X,u)|_{u=0}\gamma^\mu-\frac{i}{2}\partial^\mu_u S_{\rm th}^{\scriptscriptstyle +-}(X,u)|_{u=0}\gamma^\nu\\
&=8\pi d_q\int\frac{{\rm d}^4p}{(2\pi)^4}p^\mu p^\nu n_{\rm F}(|p^0|)\delta(p^2-g^2\sigma(X)^2)\\
&=2 d_q\int\frac{{\rm d}^3p}{(2\pi)^3}\frac{p^\mu p^\nu}{p^0} n_{\rm F}(X,{\vec p})\, .
 \end{split}
\label{eq:sc_thermTmunu}
\end{equation}
It gives the energy-momentum tensor for an ideal fluid with the energy density and the pressure obtained from the equilibrium one-loop effective potential in mean-field approximation. This is exactly what we intend to use for the fluid dynamic expansion of the quark-antiquark fluid. 

 In the present nonequilibrium model we find a correction, which from (\ref{eq:sc_approxpropexplicit}) is given by

\begin{equation}
 \partial_\mu \delta T_{\sigma}^{\mu\nu}(X)=D(X)\partial^\nu_X\sigma(X)\, .
\end{equation}
With the explicit result of the damping kernel $D(X)$ for the zero mode (\ref{eq:sc_dampingkernelapp}), the total energy-momentum dissipation from the sigma field is
\begin{equation}
 \partial_\mu T_{\sigma, {\rm appr.}}^{\mu\nu}=\left(-g\rho_s-\eta\partial_t\sigma\right)\partial^\nu\sigma\, .
\label{eq:sc_sourceterm}
\end{equation}

It includes the dissipative part of the dynamics of the sigma mean-field. It cannot, however, account for the average energy transfer from the heat bath to the field given by the auxiliary noise field $\xi$. 

What remains is the correction to the energy-momentum tensor of the quark fluid $\delta T_q^{\mu\nu}(X)$.

In upcoming works on the numerical implementation we will investigate how well the made approximations conserve energy and momentum in a fully coupled dynamic system of the chiral fields and the quark fluid.


\section{Conclusions}
In summary, we have presented a consistent nonequilibrium approach to chiral fluid dynamics, which on the one hand extends existing chiral fluid dynamic models by the inclusion of dissipation and fluctuations and on the other hand goes beyond existing studies of Langevin dynamics by putting special emphasis on the local equilibrium properties of the heat bath, i.e. on the back reaction of the chiral modes on the heat bath. 

The sigma field as the order parameter of chiral symmetry breaking is coupled to a fluid dynamic expansion of quarks and antiquarks. The interaction is given by the linear sigma model with constituent quarks, which exhibits dynamic chiral symmetry breaking. Due to this coupling the effective potential for the sigma field changes by the cooling given by the expanding quark fluid.

We succeeded in deriving the relaxational dynamics of the sigma field from the 2PI effective action. In existing chiral fluid dynamic models the sigma field is propagated according to a deterministic classical Euler-Lagrange equation of motion. The 2PI effective action includes dissipative processes and gives rise to a damping term and a stochastic field. We explicitly evaluated the damping coefficient and the correlation of the stochastic field in Markovian approximation for the zero mode of the sigma field.

Although there is no confinement in the underlying theory, the damping coefficient caused by the interaction of the sigma field with the quarks vanishes below the phase transition temperature due to kinematic reasons. While the quarks gain the constituent quark mass, the sigma mass gets smaller at the first order phase transition and very small at a critical point. Even at a realistic coupling of $g=3.3$ the vacuum sigma mass is larger than twice the constituent quark mass.

The damping coefficient can similarily be derived in the influence functional method, where an explicit splitting of the system in a relevant sector, here the sigma field, and an environment, here the quarks, must a priori be assumed. In these terms the quark fluid acts as a locally equilibrated heat bath. In the formalism of the influence funtional, however, we have no control over the equilibrium properties of the quark fluid. 

In the work presented, we put special emphasis on the consistent equilibrium properties of the heat bath. This is the main advantage of the 2PI effective action. It is a conserving and selfconsistent approximation to the full quantum theory. Besides the equation of motion for the sigma mean-field we obtain a Dyson-Schwinger equation for the real-time quark propagators. From the exact (for a given 2PI effective action) solution for the quark propagator we could construct a conserved energy-momentum tensor. For an explicit solution to the Dyson-Schwinger equation we had to make further approximations.
%
We were able to identify different terms in the divergence of the energy-momentum tensor of the entire system: a thermal part which coincides with the energy-momentum tensor for the classical fields and a correction term for both the quark and the sigma contributions to the energy-momentum balance. The correction to energy-momentum tensor of the sigma field includes the dissipative dynamics of the mean-field. However, it does not account for the fluctuation energy transferred to the heat bath via the stochastic noise field. 

The presented set up gives a consistent nonequilibrium description of the coupled dynamics of the sigma field and the quark fluid. The entire system expands and cools. It thus describes a realistic expansion of a heavy-ion collision modeled by ideal fluid dynamics. Numerical results will be published in a separate work.

Below the phase transition the zero-mode damping coefficient originating from the interaction of the sigma field with the quarks vanishes. It would be interesting to include the effect of higher modes and see how this leads to additional damping processes. However, this would go beyond the Markovian approximation and thus complicates future numerical studies. Additional damping processes potentially also come from the interaction of the soft modes of the sigma field with the hard sigma and pion modes. These processes definitely occur below the phase transition and assure relaxational dynamics of the sigma field. Especially the decay and formation processes $\sigma \leftrightarrow 2 \pi$ become important at low temperatures where the pions are light. 

The fluid dynamic treatment of the quarks might not always be valid, e.g. in the dilute phase. Starting from the Dyson-Schwinger equation one can derive a Vlasov-equation for the quark-antiquark Wigner function. In \cite{Csernai:1995zn,Mishustin:1997ff,Abada:1996bw} it was solved in the collisionless approximation. It is a more challenging task to derive dissipation and noise from a Vlasov treatment of the quarks and antiquarks and is subject to ongoing research.

\section*{Acknowledgements}
The authors thank Carsten Greiner and Igor Mishustin for fruitful and inspiring discussions. M.N. gratefully acknowledges the hospitality of the Department of Physics and Astronomy at the University of Uppsala, where main parts of this work were developed. M.N. acknowledges financial support from the Stiftung Polytechnische Gesellschaft Frankfurt am Main. This work was supported by the Hessian Initiative for Excellence (LOEWE) through the Helmholtz International Center for FAIR (HIC for FAIR).

\begin{appendix}

\section{The calculation of the influence functional}\label{sec:noneqsif}

The influence functional method gives a reduced description of the entire system with focus on the evolution of the relevant variables $\phi$. The details of the environment are eliminated by integrating out the environmental fields $\Phi$ in a path integral over the closed time path contour of the real-time description of finite temperature quantum field theory. We apply the Bogoliubov initial conditions and neglect initial correlations between the system and the environment. The interaction is then adiabatically turned on. As a consequence, the initial density matrix factorizes in system and environmental variables, $\rho_{\rm i}=\rho_{\rm i}^S\otimes\rho_{\rm i}^E$. The whole influence of the environment is then encoded in the influence functional, which is defined as

\begin{equation}\begin{split}
\exp(iS_{\rm IF}[\phi,\phi'])&=\int{\rm d}\Phi_{\rm i}\int{\rm d}\Phi_{\rm i}'\rho_{\rm i}^{\rm E}(\Phi_{\rm i},\Phi_{\rm i}')\int{\cal D}\Phi\int{\cal D}\Phi'\, {\scriptscriptstyle \times}\\
&\hphantom{=} {\scriptscriptstyle \times} \exp(iS_0[\Phi]+iS_{\rm int}[\phi,\Phi]-iS_0[\Phi']-iS_{\rm int}[\phi',\Phi'])\, .
\end{split}
\label{eq:expiSIF}
\end{equation}

Here the path integral is over all $\Phi(s)$ and $\Phi(s)'$ in real time $t_{\rm i}\leq s\leq t_{\rm f}$ with $\Phi(t_{\rm i})=\Phi_{\rm i}$ and $\Phi(t_{\rm i})=\Phi_{\rm i}'$.

In order to evaluate the explicit form of the influence functional we are often forced to make a perturbative expansion in the coupling between the two sectors. We expand the exponential function of $S_{\rm int}$ in equation (\ref{eq:expiSIF})
\begin{equation}\begin{split}
\exp(iS_{\rm IF}[\phi,\phi'])&=\int{\rm d}\Phi_{\rm i}\int{\rm d}\Phi_{\rm i}'\rho_{\rm i}^{\rm E}(\Phi_{\rm i},\Phi_{\rm i}')\int{\cal D}\Phi\int{\cal D}\Phi'\exp(iS_0[\Phi]-iS_0[\Phi']){\scriptscriptstyle \times}\\
&\hphantom{=}{\scriptscriptstyle \times}\left(1+i(S_{\rm int}[\phi,\Phi]-S_{\rm int}[\phi',\Phi'])-\frac{1}{2}(S_{\rm int}[\phi,\Phi]-S_{\rm int}[\phi',\Phi'])^2+...\right)
\end{split}
\label{eq:expiSIFexpanded}
\end{equation}

The expansion of $S_{\rm IF}[\sigma^{\scriptscriptstyle +},\sigma^{\scriptscriptstyle -}]$ becomes
\begin{equation}\begin{split}
\exp(iS_{\rm IF}[\sigma^{\scriptscriptstyle +},\sigma^{\scriptscriptstyle -}])&=\int{\rm d}\bar q^{\scriptscriptstyle +}_{\rm i}\int{\rm d}q^{\scriptscriptstyle +}_{\rm i}\int{\rm d}\bar q^{\scriptscriptstyle -}_{\rm i}\int{\rm d}q^{\scriptscriptstyle -}_{\rm i}\rho_{\rm i}^{\rm E}(\bar q^{\scriptscriptstyle +}_{\rm i},q^{\scriptscriptstyle +}_{\rm i};\bar q^{\scriptscriptstyle -}_{\rm i},q^{\scriptscriptstyle -}_{\rm i}){\scriptstyle \times}\\
&\hphantom{=}{\scriptstyle \times}\int{\cal D}\bar q^{\scriptscriptstyle +}\int{\cal D} q^{\scriptscriptstyle +}\int{\cal D}\bar q^{\scriptscriptstyle -}\int{\cal D}q^{\scriptscriptstyle -}
 \exp(iS_0[\bar q^{\scriptscriptstyle +},q^{\scriptscriptstyle +}]-iS_0[\bar q^{\scriptscriptstyle -},q^{\scriptscriptstyle -}]){\scriptstyle \times}\\
 &\hphantom{=}{\scriptstyle \times} \bigl(1- ig\int{\rm d}^4x(\bar q^{\scriptscriptstyle +}(x)q^{\scriptscriptstyle +}(x)\sigma^{\scriptscriptstyle +}(x)-\bar q^{\scriptscriptstyle -}(x)q^{\scriptscriptstyle -}(x)\sigma^{\scriptscriptstyle -}(x))\\
 &\hphantom{=\times(1}-\frac{1}{2}g^2\int{\rm d}^4x\int{\rm d}^4y
 (\bar q^{\scriptscriptstyle +}(x)q^{\scriptscriptstyle +}(x)\bar q^{\scriptscriptstyle +}(y)q^{\scriptscriptstyle +}(y)\sigma^{\scriptscriptstyle +}(x)\sigma^{\scriptscriptstyle +}(y)\\
 &\hphantom{=\times(1-\frac{1}{2}g^2\int{\rm d}^4x\int{\rm d}^4y(}-\bar q^{\scriptscriptstyle +}(x)q^{\scriptscriptstyle +}(x)\bar q^{\scriptscriptstyle -}(y)q^{\scriptscriptstyle -}(y)\sigma^{\scriptscriptstyle +}(x)\sigma^{\scriptscriptstyle -}(y)\\
 &\hphantom{=\times(1-\frac{1}{2}g^2\int{\rm d}^4x\int{\rm d}^4y(}-\bar q^{\scriptscriptstyle -}(x)q^{\scriptscriptstyle -}(x)\bar q^{\scriptscriptstyle +}(y)q^{\scriptscriptstyle +}(y)\sigma^{\scriptscriptstyle -}(x)\sigma^{\scriptscriptstyle +}(y)\\
 &\hphantom{=\times(1-\frac{1}{2}g^2\int{\rm d}^4x\int{\rm d}^4y(}+\bar q^{\scriptscriptstyle -}(x)q^{\scriptscriptstyle -}(x)\bar q^{\scriptscriptstyle -}(y)q^{\scriptscriptstyle -}(y)\sigma^{\scriptscriptstyle -}(x)\sigma^{\scriptscriptstyle -}(y)\bigr)\, .
\end{split}
\label{eq:sc_siflinsig}
\end{equation}

The definition of the free quark propagator is, for $a,b=+,\,-$,
\begin{equation}\begin{split}
iS_0^{ab}(x,y)&=\langle{\cal T}_{\cal C}q^a(x)\bar q^b(y)\rangle_0\\
	      &=\int{\rm d}\bar q^{\scriptscriptstyle +}_{\rm i}\int{\rm d}q^{\scriptscriptstyle +}_{\rm i}\int{\rm d}\bar q^{\scriptscriptstyle -}_{\rm i}\int{\rm d}q^{\scriptscriptstyle -}_{\rm i}\rho_{\rm i}^{\rm E}(\bar q^{\scriptscriptstyle +}_{\rm i},q^{\scriptscriptstyle +}_{\rm i};\bar q^{\scriptscriptstyle -}_{\rm i},q^{\scriptscriptstyle -}_{\rm i})\int{\cal D}\bar q^{\scriptscriptstyle +}\int{\cal D} q^{\scriptscriptstyle +}\int{\cal D}\bar q^{\scriptscriptstyle -}\int{\cal D}q^{\scriptscriptstyle -}{\scriptstyle \times}\\
 &\hphantom{=}{\scriptstyle \times}\exp(iS_0[\bar q^{\scriptscriptstyle +},q^{\scriptscriptstyle +}]-iS_0[\bar q^{\scriptscriptstyle -},q^{\scriptscriptstyle -}])q^a(x)\bar q^b(y)\, .
\end{split}
\label{eq:sc_defexp}
\end{equation}


For the explicit evaluation, we need the four-point functions that appear in the influence functional (\ref{eq:sc_siflinsig}). They are defined in the same way as the quark propagator (\ref{eq:sc_defexp}) and can be obtained from the generating functional 
 \begin{equation}
  Z[\bar\eta,\eta]= Z_0\exp[-\int_{\cal C}{\rm d}^4x{\rm d}^4y\bar\eta_{\cal C}(x)  S_{\cal C}(x,y)\eta_{\cal C}(y)]
 \label{eq:generatingfunctfermion}
 \end{equation}
by subsequent differentiation with respect to the external sources $\bar\eta_{\cal C}$ and $\eta_{\cal C}$. In explicit terms
\begin{subequations}
   \begin{equation}
\begin{split}
   \langle T \bar q^{\scriptscriptstyle +}(x)q^{\scriptscriptstyle +}(x)\bar q^{\scriptscriptstyle +}(y)q^{\scriptscriptstyle +}(y)\rangle&= \frac{1}{ Z_0}\left(\!\frac{i\delta}{\delta\eta^{\scriptscriptstyle +}(x)}\right)\!\!\left(\!\frac{-i\delta}{\delta\bar\eta^{\scriptscriptstyle +}(x)}\right)\!\!\left(\!\frac{i\delta}{\delta\eta^{\scriptscriptstyle +}(y)}\right)\!\!\left(\!\frac{-i\delta}{\delta\bar\eta^{\scriptscriptstyle +}(y)}\right)\! Z[\bar\eta,\eta]\bigg|_{\bar\eta=\eta=0}\\
	&=S^{\scriptscriptstyle ++}(0)^2-S^{\scriptscriptstyle ++}(x-y)S^{\scriptscriptstyle ++}(y-x)
 \end{split}
\end{equation}
\begin{equation}
 \begin{split}
   \langle T \bar q^{\scriptscriptstyle +}(x)q^{\scriptscriptstyle +}(x)\bar q^{\scriptscriptstyle -}(y)q^{\scriptscriptstyle -}(y)\rangle& = \frac{1}{ Z_0}\left(\!\frac{i\delta}{\delta\eta^{\scriptscriptstyle +}(x)}\right)\!\!\left(\!\frac{-i\delta}{\delta\bar\eta^{\scriptscriptstyle +}(x)}\right)\!\!\left(\!\frac{i\delta}{\delta\eta^{\scriptscriptstyle -}(y)}\right)\!\!\left(\!\frac{-i\delta}{\delta\bar\eta^{\scriptscriptstyle -}(y)}\right)\! Z[\bar\eta,\eta]\bigg|_{\bar\eta=\eta=0}\\
	&=S^{\scriptscriptstyle --}(0)S^{\scriptscriptstyle ++}(0)-S^{\scriptscriptstyle +-}(x-y)S^{\scriptscriptstyle -+}(y-x)
 \end{split}
\end{equation}
\begin{equation}
 \begin{split}
 \langle T \bar q^{\scriptscriptstyle -}(x)q^{\scriptscriptstyle -}(x)\bar q^{\scriptscriptstyle +}(y)q^{\scriptscriptstyle +}(y)\rangle& = \frac{1}{ Z_0}\left(\!\frac{i\delta}{\delta\eta^{\scriptscriptstyle -}(x)}\right)\!\!\left(\!\frac{-i\delta}{\delta\bar\eta^{\scriptscriptstyle -}(x)}\right)\!\!\left(\!\frac{i\delta}{\delta\eta^{\scriptscriptstyle +}(y)}\right)\!\!\left(\!\frac{-i\delta}{\delta\bar\eta^{\scriptscriptstyle +}(y)}\right)\! Z[\bar\eta,\eta]\bigg|_{\bar\eta=\eta=0}\\
	&=S^{\scriptscriptstyle ++}(0)S^{\scriptscriptstyle --}(0)-S^{\scriptscriptstyle -+}(x-y)S^{\scriptscriptstyle +-}(y-x)\\
 \end{split}
\end{equation}
\begin{equation}
 \begin{split}
\langle T \bar q^{\scriptscriptstyle -}(x)q^{\scriptscriptstyle -}(x)\bar q^{\scriptscriptstyle -}(y)q^{\scriptscriptstyle -}(y)\rangle& = \frac{1}{ Z_0}\left(\!\frac{i\delta}{\delta\eta^{\scriptscriptstyle -}(x)}\right)\!\!\left(\!\frac{-i\delta}{\delta\bar\eta^{\scriptscriptstyle -}(x)}\right)\!\!\left(\!\frac{i\delta}{\delta\eta^{\scriptscriptstyle -}(y)}\right)\!\!\left(\!\frac{-i\delta}{\delta\bar\eta^{\scriptscriptstyle -}(y)}\right)\! Z[\bar\eta,\eta]\bigg|_{\bar\eta=\eta=0}\\
	&=S^{\scriptscriptstyle --}(0)^2-S^{\scriptscriptstyle --}(x-y)S^{\scriptscriptstyle --}(y-x)\, .
\end{split}
\end{equation}
\end{subequations}

Then, neglecting all two-loop contributions, which cancel for $S^{\scriptscriptstyle ++}(0)=S^{\scriptscriptstyle --}(0)$,
\begin{equation}
 \begin{split}
  iS_{\rm IF}[\sigma^{\scriptscriptstyle+},\sigma^{\scriptscriptstyle-}] &= -\frac{1}{2}g^2\int\!\!{\rm d}^4x\int\!\!{\rm d}^4y(-S^{\scriptscriptstyle++}(x-y)S^{\scriptscriptstyle++}(y-x)\sigma^{\scriptscriptstyle+}(x)\sigma^{\scriptscriptstyle+}(y)\\
&\hphantom{= -\frac{1}{2}g^2\int\!\!{\rm d}^4x\int\!\!{\rm d}^4y(}+S^{\scriptscriptstyle+-}(x-y)S^{\scriptscriptstyle-+}(y-x)\sigma^{\scriptscriptstyle+}(x)\sigma^{\scriptscriptstyle-}(y)\\
&\hphantom{= -\frac{1}{2}g^2\int\!\!{\rm d}^4x\int\!\!{\rm d}^4y(}+S^{\scriptscriptstyle-+}(x-y)S^{\scriptscriptstyle+-}(y-x)\sigma^{\scriptscriptstyle-}(x)\sigma^{\scriptscriptstyle+}(y)\\
&\hphantom{= -\frac{1}{2}g^2\int\!\!{\rm d}^4x\int\!\!{\rm d}^4y(}-S^{\scriptscriptstyle--}(x-y)S^{\scriptscriptstyle--}(y-x)\sigma^{\scriptscriptstyle-}(x)\sigma^{\scriptscriptstyle-}(y))\, .
 \end{split}
\end{equation}
The structure of the influence functional becomes most obvious when rewriting it in terms of the center and relative field variable
\begin{subequations}
\begin{align}
 \bar\sigma&=\frac{1}{2}(\sigma^{\scriptscriptstyle +}+\sigma^{\scriptscriptstyle -})\, ,\\
 \Delta\sigma&=\sigma^{\scriptscriptstyle +}-\sigma^{\scriptscriptstyle -}\, .
\end{align}
\end{subequations}
We obtain
\begin{equation}
 \begin{split}
  iS_{\rm IF}[\bar\sigma,\Delta\sigma]&=-\frac{1}{2}g^2\int{\rm d}^4x\int{\rm d}^4y\,{\scriptstyle \times}\\
&\hphantom{=}{\scriptstyle \times}\biggl(\bar\sigma(x)\bar\sigma(y)(-S^{\scriptscriptstyle++}(x-y)S^{\scriptscriptstyle++}(y-x)+S^{\scriptscriptstyle+-}(x-y)S^{\scriptscriptstyle-+}(y-x)\\
 &\hphantom{=(\bar\sigma(x)\bar\sigma(y)\biggl(}+S^{\scriptscriptstyle-+}(x-y)S^{\scriptscriptstyle+-}(y-x)-S^{\scriptscriptstyle--}(x-y)S^{\scriptscriptstyle--}(y-x))\\
  &\hphantom{=\biggl(}+\frac{1}{2}\Delta\sigma(x)\bar\sigma(y)(-S^{\scriptscriptstyle++}(x-y)S^{\scriptscriptstyle++}(y-x)+S^{\scriptscriptstyle+-}(x-y)S^{\scriptscriptstyle-+}(y-x)\\
 &\hphantom{=+\frac{1}{2}\Delta\sigma(x)\bar\sigma(y)(}-S^{\scriptscriptstyle-+}(x-y)S^{\scriptscriptstyle+-}(y-x)+S^{\scriptscriptstyle--}(x-y)S^{\scriptscriptstyle--}(y-x))\\
  &\hphantom{=\biggl(}+\frac{1}{2}\Delta\sigma(y)\bar\sigma(x)(-S^{\scriptscriptstyle++}(x-y)S^{\scriptscriptstyle++}(y-x)-S^{\scriptscriptstyle+-}(x-y)S^{\scriptscriptstyle-+}(y-x)\\
 &\hphantom{=+\frac{1}{2}\Delta\sigma(x)\bar\sigma(y)(}+S^{\scriptscriptstyle-+}(x-y)S^{\scriptscriptstyle+-}(y-x)+S^{\scriptscriptstyle--}(x-y)S^{\scriptscriptstyle--}(y-x))\\
  &\hphantom{=\biggl(}+\frac{1}{4}\Delta\sigma(x)\Delta\sigma(y)(-S^{\scriptscriptstyle++}(x-y)S^{\scriptscriptstyle++}(y-x)-S^{\scriptscriptstyle+-}(x-y)S^{\scriptscriptstyle-+}(y-x)\\
 &\hphantom{=+\frac{1}{4}\Delta\sigma(x)\Delta\sigma(y)(}-S^{\scriptscriptstyle-+}(x-y)S^{\scriptscriptstyle+-}(y-x)-S^{\scriptscriptstyle--}(x-y)S^{\scriptscriptstyle--}(y-x))\biggl)\, .
  \end{split}
\label{eq:sc_siflinsig6}
\end{equation}
With the relations (\ref{eq:ftqft_proprel1}) and (\ref{eq:ftqft_proprel2}) the sums of products of propagators in the brackets in (\ref{eq:sc_siflinsig6}) can be evaluated. We write $S^{\scriptscriptstyle+-}=S^<$ and $S^{\scriptscriptstyle-+}=S^>$. Finally, we are left with one term that is linear and one term that is quadratic in $\Delta\sigma$,
 \begin{equation}
  \begin{split}
   iS_{\rm IF}[\bar\sigma,\Delta\sigma]
   &=-g^2\int\!\!{\rm d}^4x\int_{y_0}^{x_0}\!\!{\rm d}^4y\Delta\sigma(x)\bar\sigma(y)(S^<(x-y)S^>(y-x)-S^>(x-y)S^<(y-x))\\
    &\hphantom{=}+\frac{1}{4}g^2\int\!\!{\rm d}^4x\int\!\!{\rm d}^4y\Delta\sigma(x)\Delta\sigma(y)(S^<(x-y)S^>(y-x)+S^>(x-y)S^<(y-x))\, .
  \end{split}
 \label{eq:sc_siflinsig7}
 \end{equation}

\end{appendix}


\begin{thebibliography}{199}

\bibitem{Starobinsky:1982ee}
  A.~A.~Starobinsky,
  Phys.\ Lett.\  B {\bf 117} (1982) 175.

\bibitem{Boyanovsky:1995ema}
  D.~Boyanovsky, M.~D'Attanasio, H.~J.~de Vega, R.~Holman and D.~S.~Lee,
  Phys.\ Rev.\  D {\bf 52} (1995) 6805

\bibitem{Lacaze:2001qf}
  R.~Lacaze, P.~Lallemand, Y.~Pomeau and S.~Rica,
  Physica D {\bf 152} (2001) 779.

\bibitem{Boyanovsky:1998yp}
  D.~Boyanovsky, H.~J.~de Vega, R.~Holman and J.~Salgado,
  Phys.\ Rev.\  D {\bf 59} (1999) 125009


\bibitem{Stephanov:1998dy}
  M.~A.~Stephanov, K.~Rajagopal and E.~V.~Shuryak,
  Phys.\ Rev.\ Lett.\  {\bf 81}, 4816 (1998)

\bibitem{Stephanov:1999zu}
  M.~A.~Stephanov, K.~Rajagopal and E.~V.~Shuryak,
  Phys.\ Rev.\  D {\bf 60} (1999) 114028

\bibitem{Rybczynski:2008cv}
  M.~Rybczynski {\it et al.} [ NA49 Collaboration ],
  J.\ Phys.\ G {\bf G35 } (2008)  104091.
\bibitem{Hatta:2003wn}
  Y.~Hatta and M.~A.~Stephanov,
  Phys.\ Rev.\ Lett.\  {\bf 91} (2003) 102003
  [Erratum-ibid.\  {\bf 91} (2003) 129901]
\bibitem{Stephanov:2008qz}
  M.~A.~Stephanov,
  Phys.\ Rev.\ Lett.\  {\bf 102} (2009) 032301


\bibitem{Berdnikov:1999ph}
  B.~Berdnikov and K.~Rajagopal,
  Phys.\ Rev.\  D {\bf 61} (2000) 105017
\bibitem{Bjorken:1991xr}
  J.~D.~Bjorken,
  Int.\ J.\ Mod.\ Phys.\  A {\bf 7} (1992) 4189.

\bibitem{Rajagopal:1993ah}
  K.~Rajagopal and F.~Wilczek,
  Nucl.\ Phys.\  B {\bf 404} (1993) 577
\bibitem{Mishustin:1998eq}
  I.~N.~Mishustin,
  Phys.\ Rev.\ Lett.\  {\bf 82 } (1999)  4779-4782.
\bibitem{Randrup:1996ay}
  J.~Randrup,
  Phys.\ Rev.\ Lett.\  {\bf 77} (1996) 1226
\bibitem{Chomaz:2003dz}
  P.~Chomaz, M.~Colonna, J.~Randrup,
  Phys.\ Rept.\  {\bf 389 } (2004)  263-440.

\bibitem{Morikawa:1986rp}
  M.~Morikawa,
  Phys.\ Rev.\  {\bf D33 } (1986)  3607.

\bibitem{Gleiser:1993ea}
  M.~Gleiser and R.~O.~Ramos,
  Phys.\ Rev.\  D {\bf 50} (1994) 2441

\bibitem{Boyanovsky:1996xx}
  D.~Boyanovsky, I.~D.~Lawrie, D.~S.~Lee,
  Phys.\ Rev.\  {\bf D54 } (1996)  4013-4028.

\bibitem{Greiner:1996dx}
  C.~Greiner and B.~Muller,
  Phys.\ Rev.\  D {\bf 55} (1997) 1026

\bibitem{Bodeker:1995pp}
  D.~Bodeker, L.~D.~McLerran, A.~V.~Smilga,
  Phys.\ Rev.\  {\bf D52 } (1995)  4675-4690.

\bibitem{Son:1997qj}
  D.~T.~Son,
  [hep-ph/9707351].

\bibitem{Rischke:1998qy}
  D.~H.~Rischke,
  Phys.\ Rev.\  C {\bf 58} (1998) 2331



\bibitem{GellMann:1960np}
  M.~Gell-Mann and M.~Levy,
  Nuovo Cim.\  {\bf 16}, 705 (1960).

\bibitem{Mishustin:1998yc}
  I.~N.~Mishustin and O.~Scavenius,
  Phys.\ Rev.\ Lett.\  {\bf 83} (1999) 3134

\bibitem{Paech:2003fe}
  K.~Paech, H.~Stoecker and A.~Dumitru,
  Phys.\ Rev.\  C {\bf 68} (2003) 044907

\bibitem{Schaefer:2007pw}
  B.~J.~Schaefer, J.~M.~Pawlowski and J.~Wambach,
  Phys.\ Rev.\  D {\bf 76} (2007) 074023

\bibitem{Stokic:2008jh}
  B.~Stokic, B.~Friman, K.~Redlich,
  Phys.\ Lett.\  {\bf B673 } (2009)  192-196.


\bibitem{Scavenius:1999zc}
  O.~Scavenius and A.~Dumitru,
  Phys.\ Rev.\ Lett.\  {\bf 83} (1999) 4697

\bibitem{Luttinger:1960ua}
  J.~M.~Luttinger, J.~C.~Ward,
  Phys.\ Rev.\  {\bf 118 } (1960)  1417-1427.

\bibitem{Lee:1960zza}
  T.~D.~Lee, C.~N.~Yang,
  Phys.\ Rev.\  {\bf 117 } (1960)  22-36.

\bibitem{Baym:1961zz}
  G.~Baym, L.~P.~Kadanoff,
  Phys.\ Rev.\  {\bf 124 } (1961)  287-299.

\bibitem{Baym:1962sx}
  G.~Baym,
  Phys.\ Rev.\  {\bf 127 } (1962)  1391-1401.


\bibitem{Cornwall:1974vz}
  J.~M.~Cornwall, R.~Jackiw and E.~Tomboulis,
  Phys.\ Rev.\  D {\bf 10} (1974) 2428.

\bibitem{Ivanov:1998nv}
  Yu.~B.~Ivanov, J.~Knoll and D.~N.~Voskresensky,
  Nucl.\ Phys.\  A {\bf 657} (1999) 413

\bibitem{vanHees:2001ik}
  H.~van Hees, J.~Knoll,
  Phys.\ Rev.\  {\bf D65 } (2002)  025010.

\bibitem{vanHees:2001pf}
  H.~van Hees, J.~Knoll,
  Phys.\ Rev.\  {\bf D65 } (2002)  105005.

\bibitem{vanHees:2002bv}
  H.~van Hees, J.~Knoll,
  Phys.\ Rev.\  {\bf D66 } (2002)  025028.


\bibitem{Jungnickel:1995fp}
  D.~U.~Jungnickel, C.~Wetterich,
  Phys.\ Rev.\  {\bf D53 } (1996)  5142-5175.

\bibitem{Berges:1998ha}
  J.~Berges, D.~U.~Jungnickel, C.~Wetterich,
  Int.\ J.\ Mod.\ Phys.\  {\bf A18 } (2003)  3189-3220.

\bibitem{Tetradis:2003qa}
  N.~Tetradis,
  Nucl.\ Phys.\  {\bf A726 } (2003)  93-119.

\bibitem{Schaefer:2006sr}
  B.~-J.~Schaefer, J.~Wambach,
  Phys.\ Part.\ Nucl.\  {\bf 39 } (2008)  1025-1032.


\bibitem{Skokov:2010sf}
  V.~Skokov, B.~Friman, E.~Nakano, K.~Redlich, B.~-J.~Schaefer,
  Phys.\ Rev.\  {\bf D82 } (2010)  034029.


\bibitem{Mocsy:2004ab}
  A.~Mocsy, I.~N.~Mishustin, P.~J.~Ellis,
  Phys.\ Rev.\  {\bf C70 } (2004)  015204.

\bibitem{Palhares:2008yq}
  L.~F.~Palhares, E.~S.~Fraga,
  Phys.\ Rev.\  {\bf D78 } (2008)  025013.

\bibitem{Fraga:2009pi}
  E.~S.~Fraga, L.~F.~Palhares, M.~B.~Pinto,
  Phys.\ Rev.\  {\bf D79 } (2009)  065026.

\bibitem{Palhares:2010be}
  L.~F.~Palhares, E.~S.~Fraga,
  Phys.\ Rev.\  {\bf D82 } (2010)  125018.

\bibitem{Scavenius:2000qd}
  O.~Scavenius, A.~Mocsy, I.~N.~Mishustin and D.~H.~Rischke,
  Phys.\ Rev.\  C {\bf 64} (2001) 045202

\bibitem{Aoki:2006we}
  Y.~Aoki, G.~Endrodi, Z.~Fodor, S.~D.~Katz, K.~K.~Szabo,
  Nature {\bf 443 } (2006)  675-678.

\bibitem{Friman:2011zz}
  B.~Friman, C.~H\"ohne, J.~Knoll, S.~Leupold, J.~Randrup, R.~Rapp, P.~Senger (eds.),
  Lect.\ Notes Phys.\  {\bf 814 } (2011)  1-980. 
\bibitem{Aguiar:2003pp}
  C.~E.~Aguiar, E.~S.~Fraga and T.~Kodama,
  J.\ Phys.\ G {\bf 32} (2006) 179

 \bibitem{Scavenius:2000bb}
   O.~Scavenius, A.~Dumitru, E.~S.~Fraga, J.~T.~Lenaghan, A.~D.~Jackson,
   Phys.\ Rev.\  {\bf D63 } (2001)  116003.

\bibitem{Feynman:1963fq}
  R.~P.~Feynman and F.~L.~.~Vernon,
  Annals Phys.\  {\bf 24} (1963) 118
  [Annals Phys.\  {\bf 281} (2000) 547].

\bibitem{Greiner:1998vd}
  C.~Greiner and S.~Leupold,
  Annals Phys.\  {\bf 270} (1998) 328

\bibitem{Schwinger:1960qe}
  J.~S.~Schwinger,
  J.\ Math.\ Phys.\  {\bf 2} (1961) 407.

\bibitem{Keldysh:1964ud}
  L.~V.~Keldysh,
  Zh.\ Eksp.\ Teor.\ Fiz.\  {\bf 47} (1964) 1515
  [Sov.\ Phys.\ JETP {\bf 20} (1965) 1018].

\bibitem{Hu:1991di}
  B.~L.~Hu, J.~P.~Paz, Y.~Zhang,
  Phys.\ Rev.\  {\bf D45 } (1992)  2843-2861.

\bibitem{Hu:1993vs}
  B.~L.~Hu, J.~P.~Paz, Y.~Zhang,
  Phys.\ Rev.\  {\bf D47 } (1993)  1576-1594.

\bibitem{GellMann:1992kh}
  M.~Gell-Mann, J.~B.~Hartle,
  Phys.\ Rev.\  {\bf D47 } (1993)  3345-3382.

\bibitem{Calzetta:1995ys}
  E.~Calzetta, B.~L.~Hu,
  Phys.\ Rev.\  {\bf D52 } (1995)  6770-6788.

\bibitem{Calzetta:1995ea}
  E.~Calzetta, B.~L.~Hu,
    [hep-th/9501040].

\bibitem{einstein}
  A.~Einstein, Annalen der Physik {\bf 322} (1905)  549


\bibitem{Leupold:2006bp}
  S.~Leupold,
  Phys.\ Lett.\  B {\bf 646} (2007) 155

\bibitem{Berges:2001fi}
  J.~Berges,
  Nucl.\ Phys.\  {\bf A699 } (2002)  847-886.

\bibitem{Juchem:2003bi}
  S.~Juchem, W.~Cassing, C.~Greiner,
  Phys.\ Rev.\  {\bf D69 } (2004)  025006.

\bibitem{Juchem:2004cs}
  S.~Juchem, W.~Cassing, C.~Greiner,
  Nucl.\ Phys.\  {\bf A743 } (2004)  92-126.

\bibitem{Kap94}
 J.~I.~Kapusta, Finite-Temperature Field Theory, (University Press, Cambridge, 1994)

\bibitem{Bog62}
N. N. Bogoliubov, in ``Studies in Statistical Mechanics``, edited by I. de Boer and G. E. Uhlenbeck (North-Holland, Amsterdam, 1962), Vol. I;

\bibitem{Das97}
  A.~Das, Finite Temperature Field Theory, (World Scientific Publishing Company, 1997)
\bibitem{Landsman:1986uw}
  N.~P.~Landsman and C.~G.~van Weert,
  Phys.\ Rept.\  {\bf 145} (1987) 141.
%
%

\bibitem{Weldon:1983jn}
  H.~A.~Weldon,
  Phys.\ Rev.\  D {\bf 28} (1983) 2007.

\bibitem{Biro:1997va}
  T.~S.~Biro and C.~Greiner,
  Phys.\ Rev.\ Lett.\  {\bf 79} (1997) 3138

\bibitem{Xu:1999aq}
  Z.~Xu and C.~Greiner,
  Phys.\ Rev.\  D {\bf 62} (2000) 036012

\bibitem{Csernai:1995zn}
  L.~P.~Csernai, I.~N.~Mishustin,
  Phys.\ Rev.\ Lett.\  {\bf 74 } (1995)  5005-5008.
\bibitem{Mishustin:1997ff}
  I.~N.~Mishustin, O.~Scavenius,
  Phys.\ Lett.\  {\bf B396 } (1997)  33-38.

\bibitem{Abada:1996bw}
  A.~Abada, M.~C.~Birse,
  Phys.\ Rev.\  {\bf D55 } (1997)  6887-6899.







%

%
%
%
%
%




\end{thebibliography}
\end{document}